\documentclass[a4paper,useAMS,fleqn,usenatbib]{mnras}

\usepackage{graphicx}
\usepackage{natbib}
\usepackage{amsmath}
\usepackage{mathtools}
\usepackage{amssymb}
\usepackage{wasysym}
\usepackage{txfonts}
\usepackage[T1]{fontenc}
\usepackage{ae,aecompl}

\usepackage{latexsym,longtable,epsf,ulem}
\usepackage{cases} 
\usepackage{enumerate} 
\usepackage{ulem}

\newcommand{\ant}{\alpha_{\rm nt}}
\newcommand{\fth}{f_{\rm th}}

\title[Dwarf galaxy: IC\,10]
{New insights into the interstellar medium of the dwarf galaxy IC\,10: connection between
magnetic fields, the radio--infrared correlation and star formation}
\author[Basu et al.]{Aritra Basu$^1$\thanks{E-mail : abasu@mpifr-bonn.mpg.de (AB); sambit.roychowdhury@manchester.ac.uk (SR)}, Sambit Roychowdhury$^{2\star}$, Volker Heesen$^3$, Rainer Beck$^1$, Elias Brinks$^4$, 
\newauthor Jonathan Westcott$^4$ and Luke Hindson$^4$
	\\
$^1$Max-Planck-Institut f{\"u}r Radioastronomie, Auf dem H{\"u}gel 69, D-53121 Bonn, Germany\\
$^2$Jodrell Bank Centre for Astrophysics, Alan Turing Building, School of Physics \& Astronomy, The University of Manchester, \\Oxford Road, Manchester M13 9PL, UK\\
$^3$Universit{\"a}t Hamburg, Hamburger Sternwarte, Gojenbergsweg 112, D-21029 Hamburg, Germany\\
$^4$Centre for Astrophysics Research, University of Hertfordshire, Hatfield AL10 9AB, UK}

\begin{document}

\date{\it Accepted to be published in MNRAS: 2017 June 20}

\pagerange{\pageref{firstpage}--\pageref{lastpage}} \pubyear{2002}

\maketitle

\label{firstpage}

\begin{abstract}

We present the highest sensitivity and angular resolution study at 0.32 GHz of
the dwarf irregular galaxy IC\,10, observed using the Giant Metrewave Radio
Telescope, probing $\sim45$ pc spatial scales. We find the galaxy-averaged
radio continuum spectrum to be relatively flat, with a spectral index $\alpha =
-0.34\pm0.01$ ($S_\nu \propto \nu^\alpha$), mainly due to a high contribution
from free--free emission. At 0.32 GHz, some of the H{\sc ii} regions show
evidence of free--free absorption as they become optically thick below
$\sim0.41$ GHz with corresponding free electron densities of $\sim11-22~\rm
cm^{-3}$.  After removing the free--free emission, we studied the
radio--infrared relations on 55, 110 and 165 pc spatial scales. We find that on
all scales the non-thermal emission at 0.32 and 6.2 GHz correlates better with
far-infrared (FIR) emission at $70\,\mu$m than mid-infrared emission at
$24\,\mu$m.  The dispersion of the radio--FIR relation arises due to variations
in both magnetic field and dust temperature, and decreases systematically with
increasing spatial scale. The effect of cosmic ray transport is negligible as
cosmic ray electrons were only injected $\lesssim5$ Myr ago. The average
magnetic field strength ($B$) of $12~\mu$G in the disc is comparable to that of
large star-forming galaxies. The local magnetic field is strongly correlated
with local star formation rate ($\mathrm{SFR}$) as $B \propto
\mathrm{SFR}^{0.35\pm0.03}$, indicating a star-burst driven fluctuation dynamo
to be efficient ($\sim10$ per cent) in amplifying the field in IC\,10. The high
spatial resolution observations presented here suggest that the high efficiency
of magnetic field amplification and strong coupling with SFR likely sets up the
radio--FIR correlation in cosmologically young galaxies.

\end{abstract}


\section{Introduction}

According to models of hierarchical structure formation, low mass and low
luminosity dwarf irregular galaxies are thought to be the analogues of the
first galaxies that formed in the early Universe which evolved into larger
systems like the normal star-forming spirals found in the Local Volume.  Dwarf
irregular galaxies differ from normal star-forming galaxies in terms of their
global properties, such as size, structure and heavy element abundance
(metallicity) of the interstellar medium (ISM) which follows the
mass-metallicity relation \citep{skill89, riche95, berg12} and is typically in
the range 0.1--0.3 $Z_\odot$.  These galaxies have low stellar mass because of
their small sizes, but can have large gas-to-stellar mass ratio compared to
that in spiral galaxies \citep{begum05, ott12, hunter12, mcnic16}. As a
consequence of their low mass, their rotational velocities are low
\citep{broei97, begum08, oh08, ott12, mcnic16}. Because the velocity dispersion
of the gas is similar to that in large spirals, this implies that their ISM
forms a thick disc with scale heights of several hundred parsecs
\citep{baner11}. Dwarf galaxies lack any coherent structures such as spiral
arms or central bars, and the structure of their discs is different from that
in normal spirals \citep[e.g.,][]{heidm72, stave92, sanch10, roych10}. Thus,
dwarf galaxies are fundamentally different in terms of the physical nature of
the ISM as compared to that of large star-forming galaxies.  Studies of these
objects may provide important clues linked to the cosmic evolution of ISM
properties in normal galaxies.

Star-forming dwarf galaxies are believed to dominate the population of galaxies
that formed earliest and which contributes significantly to the cosmic star
formation rate around redshift $\sim2$ \citep{kanek09, kanek14, buitr13,
alavi16}. However, dwarf galaxies are significantly fainter
($M_B\gtrsim -19$) than large spirals ($M_B \lesssim -20$), and studying star
formation at the faint end of the galaxy luminosity function is challenging,
especially in the early universe \citep{jarvi15}. Of late, by taking advantage
of the radio--far infrared (FIR) relation, star formation in the early universe
is being indirectly traced by its associated radio continuum emission.
The radio--FIR relation alludes to the correlation between non-thermal radio
(primarily at 1.4 GHz) and FIR (in one band or bolometric) luminosities over at
least 4 orders of magnitude and across galaxy types \citep{wunde87, dress88,
condo92, price92, yun01}.

Since the discovery of the correlation, several models have been proposed to
explain its origin \citep[see e.g.,][]{volk89, helou93, nikla97b, bell03,
lacki10}, however, a clear understanding of the reason behind the correlation,
across galaxies with a range of star formation properties, magnetic field
strengths, etc., remains elusive. It is believed that star formation plays a
pivotal role for the correlation, as it is directly responsible for the thermal
part of the radio emission, and indirectly responsible for the non-thermal part
of the radio emission and the re-radiated emission by dust in the infrared
band.  Of late, a framework based on theoretical and empirical results on
amplification of magnetic fields in galaxies and how it is coupled with the gas
density has emerged, which may well be able to fully explain the existence of
the radio--FIR relation in galaxies \citep{schle13, schle16, schob16}. At its
root lies magnetohydrodynamic (MHD) turbulence, which efficiently amplifies the
magnetic fields on small-scales ($\lesssim1$ kpc) to energy equipartition
values and drives the coupling between magnetic field and gas density which
helps in maintaining the tightness of the relation.

Since the gas motions in the ISM of irregular galaxies are predominantly driven
by turbulent motions, maintained by kinetic energy input from supernovae
explosions, the fluctuation dynamo mechanism (which must be prevalent in them)
is much more efficient than the classical $\alpha$--$\Omega$ dynamo, the
fields are expected to be in energy equipartition. Therefore, we put forth the
conjecture that observations of the radio--FIR relation on sub-kpc scales in
dwarf irregular galaxies can act as a test of the theoretical framework
mentioned above, for explaining the correlation. Historically, the
relation has been shown to exist for galaxy-averaged luminosities
\citep{condo92, yun01}, but recent studies of spatially resolved regions in
galaxies show the correlation to hold with differences in slope dependent on
local features and whether or not energy `equipartition' conditions are valid
\citep{hoern98, murgi05, hughe06, dumas11, basu12a}. The correlation has also
been shown to exist for the faintest star forming dwarf galaxies
(\citealt{hughe06, roych12}; Kitchener et al. 2017, AJ, submitted).

\begin{table} \centering 
 \caption{Properties of IC\,10 and sources of data used.} 
  \begin{tabular}{@{}lc@{}} 
 \hline 
Galaxy type & IB \\
Angular size (D$_{25}$) & $6.8^\prime\times5.9^\prime$ \\ 
Inclination$^\dagger$ & 31 degrees \\
Distance$^1$ & 0.74 Mpc \\
SFR$^2$ & 0.05--0.2 $\rm M_\odot\,yr^{-1}$\\
Dynamical mass$^2$ & $1.7\times10^9$ $\rm M_\odot$ \\
\hline
\hline
Radio continuum data: &\\
0.32 GHz & GMRT \\
1.4 GHz & VLA (C-array)$^a$\\
6.2 GHz & VLA+Effelsberg$^b$\\
 & \\
H{\sc i} & VLA (LITTLE THINGS)\\
 & \\
Infrared data: &\\
$24\,\mu$m & {\it Spitzer} MIPS\\
$70\,\mu$m & {\it Herschel} PACS\\
$160\,\mu$m & {\it Herschel} PACS\\
 & \\
H$\alpha$ &  Perkins 1.8-m (Lowell observatory) \\

\hline 
\end{tabular}
\begin{flushleft}
$^\dagger$The inclination angle ($i = 0^\circ$ is face-on) is taken from
HyperLEDA. $^1$ The distance is adopted from \cite{tully13}. $^2$ The
star-formation rate (SFR) and the dynamical mass is taken from the compilation
of \citet{leroy06} and the references therein. \\
$^a$ Archival VLA data observed in 2004 (project code: AC717).\\
$^b$ Image from \citet{heese15}.
\end{flushleft}
\label{sampletab} 
\end{table}

In the radio continuum, the brighter end of the galaxy luminosity function has
been studied in detail, however such studies have been lacking for irregular
star-forming dwarf galaxies, especially those that provide spatially resolved
studies of the radio emission. At a distance of 0.74 Mpc, IC\,10 is the nearest
metal-poor dwarf irregular galaxy which has undergone a star-burst phase
$\sim10$ Myr ago \citep{vacca07}. Thus, IC\,10 is a prototypical example
of a cosmologically young galaxy in the nearby universe. This galaxy has been
studied extensively from radio to X-rays through infrared (IR) and optical
wavebands.  Because of its proximity, IC\,10 is bright in all  wavebands,
making it an ideal candidate to perform spatially resolved studies. The
properties of IC\,10 and the data used in this work and their provenance, are
listed in Table~\ref{sampletab}.

In this paper, we present the lowest radio-frequency observations of IC\,10 at
0.32 GHz published to date. Using these observations and archival data at
higher frequencies, we study the spatially resolved radio continuum spectra of
IC\,10 and the radio--infrared relation at $\lesssim200$ pc spatial scales.
We summarize existing radio continuum studies of IC\,10 in Section 2. In
Section 3 we present the data analysis procedure. The results of the new
low-frequency observations, magnetic field strengths and radio--infrared
relation are presented in Section 4. In Section 5 we discuss our results and
summarize them in Section 6.

\section{Radio continuum studies of IC\,10} \label{studies}

The dwarf galaxy IC\,10 has been the subject of several radio continuum
studies. \citet{klein86} performed single dish observations of IC\,10 and found
a significantly flatter radio continuum spectrum between 1 and 10 GHz with
spectral index, $\alpha = -0.33 \pm 0.03$ (defined as $S_\nu \propto
\nu^\alpha$) as compared to normal star-forming galaxies which typically have
$\alpha$ in the range $-0.6$ and $-1.0$. In these low resolution observations,
emission coincident with IC\,10's bright H{\sc ii} regions was detected. High
resolution interferometric observations at 6 GHz by \citet{heese11} revealed
the radio continuum emission to trace the H$\alpha$-emitting disc of IC\,10.
They estimated that $\sim50$ per cent of the radio continuum emission at 6 GHz
arises due to thermal free--free emission which makes it a good tracer of star
formation. \citet{westc17} observed IC\,10 at sub-arcsec angular resolutions at
1.5 and 5 GHz. They detected 11 compact sources within the disc of IC\,10 and
identified 5 sources as background sources and 3 sources as compact H{\sc ii}
regions that are coincident with peaks in H$\alpha$ emission.

\citet{yang93}, in their observations at 0.608, 1.464 and 4.87 GHz, discovered
the presence of a non-thermal superbubble towards the south-eastern edge of the
galaxy, likely the result of several supernovae. In a detailed study of the
superbubble between 1.5 and 8.8 GHz, \citet{heese15} found evidence of its
non-thermal radio continuum spectrum being curved, which steepens
towards higher frequencies, which is caused by energy loss of the synchrotron
emitting cosmic ray electrons (CREs) accelerated in magnetic field of
$44~\mu$G. Expansion of the superbubble produced enhanced ordered magnetic
fields giving rise to strong polarized emission \citep{heese11}.

\citet{chyzy16} performed deep observations of IC\,10 at 1.43 GHz.  These
observations show the likely existence of a spherical-shaped radio continuum
halo extending up to $\sim 2$ kpc, i.e., about a factor of 2 larger than the
H$\alpha$-emitting disc. The halo hosts an X-shaped magnetic field structure,
which they argue is produced by a large-scale magnetized galactic wind driven
by star formation. Overall, they find the total magnetic field strength to be
stronger than in other dwarf galaxies, while the galaxy
averaged emission follows the global radio--FIR relation.

Although several detailed studies of special features in IC\,10 exist in the
literature, spatially resolved studies of star-forming dwarf galaxies are
generally lacking in the literature. Also, since there is a high contribution of
thermal free--free emission to the total radio continuum emission in IC\,10,
low radio-frequency ($<1$ GHz) observations are necessary to study the
non-thermal emission -- thus motivating this investigation.

\begin{figure}
\begin{centering}
\begin{tabular}{c}
{\mbox{\includegraphics[width=9cm, trim=0mm 5mm 0mm 5mm,clip]{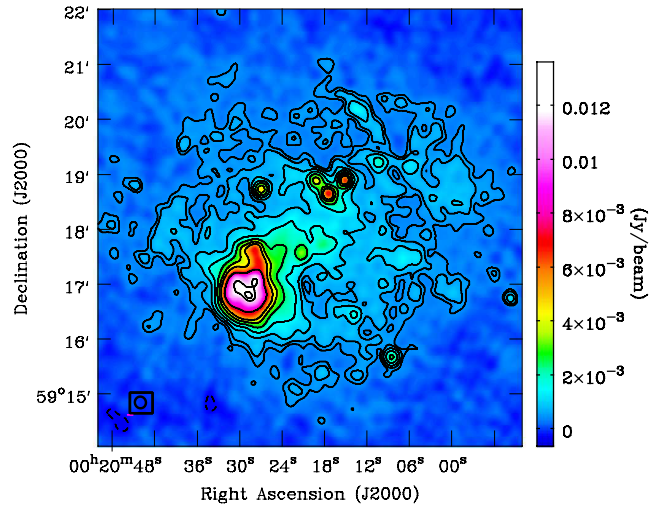}}} \\
\end{tabular}
\end{centering}
\caption{Total intensity map of IC\,10 at 0.32 GHz observed using the GMRT with
an angular resolution of $13\times12$ arcsec$^2$. The beam is shown in bottom
left hand corner. The contour levels are at $(-3, -2, 3, 4, 6, 10, 16, 24, 32,
50, 80, 85)\times 150~\mu$Jy beam$^{-1}$.}
\label{totI_ic10}
\end{figure}

\section{Observations and data analysis}

\subsection{GMRT data}

We observed IC\,10 using the Giant Metrewave Radio Telescope (GMRT) at 0.322
GHz ($\nu_{\rm RF}$ between 0.307 to 0.339 GHz) using the software backend
having a bandwidth of 33.3 MHz split into 512 channels and a total on-source
time of $\sim7$ hours. The data was analyzed using the NRAO Astronomical Image
Processing System\footnote{AIPS is produced and maintained by the National
Radio Astronomy Observatory, a facility of the National Science Foundation
operated under cooperative agreement by Associated Universities, Inc.} (AIPS)
following standard routines for data flagging and calibration. The phase
solutions were iteratively obtained on a nearby bright {\it phase calibrator},
3C\,468.1. The solutions were transferred to the target source once the closure
errors were less than 1 per cent. The phase calibrator 3C\,468.1 has a flux
density of 20.5 Jy at 0.32 GHz and was used to solve for the bandpass gains
along with the {\it flux calibrator} 3C\,147. 

\begin{figure}
\begin{centering}
\begin{tabular}{c}
{\mbox{\includegraphics[width=8cm,trim=0mm 0mm 0mm 0mm, clip]{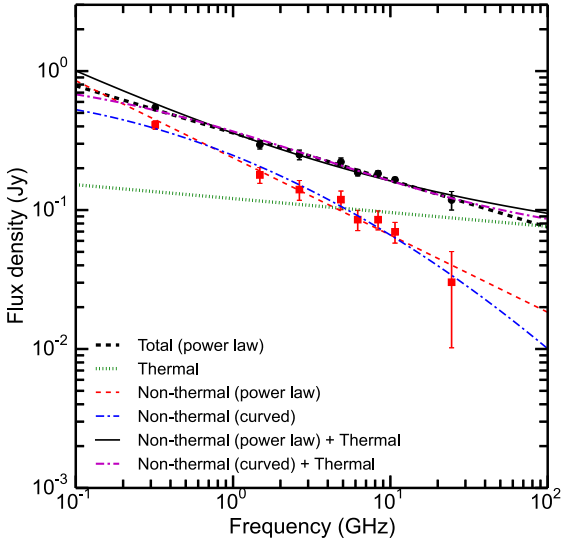}}}
\end{tabular}
\end{centering}
\caption{Galaxy-integrated radio continuum spectrum of IC\,10. The black data
points are the flux densities measured at various frequencies (see
Table~\ref{tableflux}). The black dashed line shows the classical power-law fit to
the data having spectral index $-0.34\pm0.01$.  The green dotted line is the
estimated thermal emission.  The red points show the non-thermal flux
densities after subtracting the thermal emission and the red line is the
best-fit power-law with $\ant = -0.55\pm0.04$. The blue dot-dashed line is the
best-fit to the non-thermal emission with a curved spectrum. The solid black and
dot-dashed magenta curves are the total intensity spectrum estimated by adding
the thermal emission to the non-thermal emission estimated by power-law
and curved-spectrum, respectively.}
\label{integflux}
\end{figure}

Several rounds of {\it phase-only} self-calibration were performed using the
point sources in the target field. To achieve a reliable model of the point
sources, we used only the baselines $\gtrsim1$ k$\lambda$ and the data was
weighted using Briggs' robust parameter of $-1$ \citep{brigg95}.  Final imaging
was done using all baselines and employing the technique of polyhedron imaging
for making wide-field images with non-coplanar baselines.  To deconvolve the
diffuse emission, we employed the {\sc sdi-clean} algorithm \citep{steer84} in
AIPS and used a Briggs' robust parameter of 0. In Fig.~\ref{totI_ic10}, we show
the total intensity image of IC\,10 at 0.32 GHz. The resolution of the
image\footnote{This corresponds to a projected linear resolution of $\sim46.5$
pc at a distance of 0.74 Mpc.}  is $13\times12$ arcsec$^{2}$ and has rms
noise of $150~\mu$Jy beam$^{-1}$. The galaxy-integrated flux density is found
to be 545$\pm$25 mJy and is in agreement with extrapolated flux densities from
higher frequencies (see Fig.~\ref{integflux}).  Note that the largest angular
scale detectable at 0.32 GHz by the GMRT is $\sim18$ arcmin. Therefore, at the
short baselines ($\rm \lesssim 0.2~k\lambda$), IC\,10 is unresolved and hence
we do not expect any missing flux-density of the diffuse emission in our
observations and are only noise-limited.

\subsection{Archival data at higher radio frequencies}

\subsubsection{1.42 GHz} \label{lband}

We downloaded and analyzed archival dataset of IC\,10 observed in 2004
with the NRAO\footnote{The NRAO is a facility of the National Science
Foundation operated under cooperative agreement by Associated Universities,
Inc.} Karl G.\ Jansky Very Large Array (VLA) in C-configuration using two
side-bands of 50 MHz each, centered at 1.385 and 1.465 GHz (project code:
AC717). The total on-source time was $\sim5$ hours. Combining the two
side-bands we obtain a map of IC\,10 with 100 MHz bandwidth centered at 1.42
GHz. After standard data analysis we obtain a map rms noise of $\sim35~\mu$Jy
beam$^{-1}$ and an angular resolution of $17\times16$ arscec$^2$.

From these observations, the total flux density of IC\,10 at 1.42 GHz is found
to be $270\pm20$ mJy.  At similar frequencies, the flux density measured by the
NVSS data \citep{condo98} is $300\pm20$ mJy whereas \citet{chyzy16} found a
total flux density of $377\pm11$ mJy, which includes a large extended
synchrotron halo. We therefore believe that the archival C-configuration data
suffers from missing flux density at $\gtrsim20$ per cent level. However, the
missing large angular scale emission does not affect the flux densities at
small-scales, especially for sources that are unresolved. We therefore
restricted the use of these data to study the spectrum of compact regions
within the disc of IC\,10.

\subsubsection{6.2 GHz}

IC 10 was observed using the VLA in D-configuration at a central frequency of
6.2 GHz \citep[project code: AH1006;][]{heese15}. To mitigate the effects
of missing large-angular scale emission in these interferometric observations,
single-dish data observed using the Effelsberg 100-m telescope were
added to obtain total intensity image at $9.4\times7.3$ arcsec$^2$ angular
resolution and rms noise of $15~\mu$Jy beam$^{-1}$ \citep[see][for
details]{heese15}. The total flux density of the combined VLA+Effelsberg
observations is found to be $186\pm10$ mJy at 6.2 GHz.

\subsection{Other ancillary data}

\subsubsection{H$\alpha$ map}

We used H$\alpha$ emission of IC\,10 to estimate the contribution of
thermal free--free emission to the total radio continuum emission and to
compute the star formation rate (SFR). A stellar-continuum subtracted H$\alpha$
image, observed using the Perkins 1.8-m telescope at the Lowell observatory
\citep{hunter04}, was downloaded from the NED. The image has an angular
resolution of $2.2 \times 2.2$ arcsec$^2$ and a brightness sensitivity of
$10^{-18}~\rm erg\,s^{-1}\,cm^{-2}$ per 0.49 arcsec pixel-size. This
corresponds to a star formation rate surface density of $1.1\times 10^{-4}~\rm
M_\odot\,yr^{-1}\,kpc^{-2}$ (not corrected for extinction) at distance of 0.74
Mpc for IC\,10.

\subsubsection{Infrared maps}

{\it Spitzer MIPS 24\,$\mu$m}: To correct for internal dust extinction of the
H$\alpha$ emission and to study the spatially resolved radio--IR relation, we used
a $24\,\mu$m map observed using the {\it Spitzer} space telescope
\citep{bendo12}. The map has an angular resolution of $6 \times 6$ arcsec$^2$
and a $1\sigma$ flux density sensitivity of $0.25$ mJy beam$^{-1}$.
\\
{\it Herschel PACS 70\,$\mu$m}: To study the radio--IR relation using cold dust
emission and estimate its temperature we used a far-infrared map of IC\,10 at
$70\,\mu$m. IC\,10 was observed using the {\it Herschel} space telescope and we
downloaded a level 2.5 data product from the NASA/IPAC Infrared Science Archive
(IRSA).\footnote{http://irsa.ipac.caltech.edu/applications/Herschel/} The
$70\,\mu$m-map has an angular resolution of $5.2 \times 5.2$ arcsec$^2$ with a
$1\sigma$ flux density sensitivity of $\sim6$ mJy beam$^{-1}$.
\\
{\it Herschel PACS 160\,$\mu$m}: For computing the dust temperature, in
addition to the $70\,\mu$m map of IC\,10, we downloaded a $160\,\mu$m-map
observed using {\it Herschel} from the IRSA. The map was available with $12
\times 12$ arcsec$^2$ angular resolution and a $1\sigma$ rms noise of $\sim80$
mJy beam$^{-1}$.

\subsection{Thermal emission separation} \label{thermalsep}

Since IC\,10 has recently undergone an active burst of star formation
\citep[][]{vacca07, yin10}, the contribution of the thermal free--free emission
to the total radio continuum emission could be significant. We estimated the
thermal emission from IC\,10 using the H$\alpha$ emission as its tracer,
because it originates from the recombination of the same free electrons that
produces the free--free emission. The thermal flux density ($S_{\rm th,\nu}$) at
a radio frequency $\nu$ is related to the electron temperature ($T_{\rm e}$,
assumed to be $10^4$ K) and the free--free optical depth ($\tau_{\rm ff}$) as,
\begin{equation}
S_{\rm th, \nu} = \frac{2\, k\, T_{\rm e} \nu^2}{c^2} (1 - e^{-\tau_{\rm ff}}). \label{eqn_sth}
\end{equation}
$\tau_{\rm ff}$ is related to the emission measure (EM) as,
\begin{equation}
\tau_{\rm ff} = 0.082\, T_{\rm e}^{-1.35} \left(\frac{\nu}{\rm GHz} \right)^{-2.1} \left(\frac{\rm EM}{\rm cm^{-6} pc} \right), \label{tauff}
\end{equation}
where the EM is determined from the H$\alpha$ intensity ($I_{\rm H\alpha}$)
following \citet{valls98},
\begin{equation}
\left(\frac{I_{\rm H\alpha}}{\rm erg\,cm^{-2}s^{-1}sr^{-1}}\right) = 9.41\times 10^{-8} T_{\rm e4}^{-1.017} 10^{-0.029/T_{\rm e4}} \left(\frac{\rm EM}{\rm cm^{-6} pc} \right).
\end{equation}
Here, $T_{\rm e4}$ is the electron temperature in units of $10^4$ K and, $k$
and $c$ are the standard constants.

Although the H$\alpha$ emission is the best tracer of the thermal emission, it
is easily absorbed by the dust present both internal to the galaxy and in the
Galactic foreground (Milky Way). The Galactic latitude of IC\,10 is
$-3.3^\circ$, and therefore Milky Way extinction is significant. The
standard \citet{schlegel98} maps for estimating the Milky Way dust extinction
are not useful below Galactic latitude $\sim5^{\circ}$.  Instead, we use
the average value of $E(B-V)=0.77$ measured by \citet{riche01} in the direction
of IC\,10, which combined with an $R_V$ of $3.1$ gives the value for extinction
at the wavelength of H$\alpha$ to be $A_{H \alpha}=1.95$. We apply this as
the uniform Galactic extinction for IC\,10. On the other hand, the internal
extinction of the H$\alpha$ emission was accounted for by combining the
extinction-corrected H$\alpha$ emission with 24-$\mu$m emission of IC\,10 as
\citep{kenni09},
\begin{equation}
I_{\rm H\alpha, corr} = I_{\rm H\alpha,obs} 10^{A_{\rm H\alpha}/2.5} + 0.02\, \nu_{\rm 24 \mu m}\, I_{\rm 24\mu m}. \label{eqn_hacorr}
\end{equation}
Here, $I_{\rm H\alpha,obs}$ and $I_{\rm H\alpha, corr}$ are the observed
and extinction-corrected H$\alpha$ intensities, respectively and $I_{\rm 24\mu
m}$ is the intensity of the $24\,\mu$m emission. Due to the low metallicity of
IC\,10 ($Z\approx 0.2\, Z_\odot$, \citealt{garne90}), overall, the internal
extinction is found to be $\lesssim30$ per cent. Locally, the internal
extinction can be up to 60 per cent in bright H{\sc ii} regions while in the
disc it lies in the range 5--20 per cent.

Both the H$\alpha$ and the $24\,\mu$m maps were convolved to a common angular
resolution of 15 arcsec, determined by the radio continuum maps. They were then
aligned to a common coordinate system.  After correcting for dust extinction of
the H$\alpha$ emission, the thermal emission was estimated on a pixel-by-pixel
basis and subtracted from the total radio continuum emission at 0.32, 1.42 and
6.2 GHz.  Overall, we estimate the thermal fraction\footnote{The thermal
fraction at a frequency $\nu$ is defined as $f_{\rm th, \nu} = S_{\rm th,
\nu}/S_\nu$. Here, $S_\nu$ is the total radio continuum emission.}, $\fth$, to
be $0.20\pm0.05$, $0.35\pm0.07$ and $0.53\pm0.1$ at 0.32, 1.42 and 6.2 GHz,
respectively. As mentioned in Section~\ref{lband}, the 1.42-GHz radio continuum map
likely suffers from missing flux density and hence, $f_{\rm th, 1.42\,GHz}$ is
overestimated. However, this is not an issue at 0.32 and 6.2 GHz. 

Uncertainty in the estimated thermal emission mainly arises from the
extinction correction of the H$\alpha$ emission and from $T_{\rm e}$,
which is not well known. \citet{kenni09} pointed out that the combination of
H$\alpha$ and $24\,\mu$m emission suffers small systematic variation compared
to reliable tracers of extinction such as the Balmer decrement and are within
$\sim15$ per cent on average. Moreover, because of the low metallicity of
IC\,10, the internal extinction is low and hence we do not expect significant
uncertainty due to the internal attenuation correction. For example, assuming a
30 per cent uncertainty due to the internal extinction applied through
Eq.~\ref{eqn_hacorr}, the maximum uncertainty to the thermal emission is
$\sim10$ per cent and is $\lesssim 5$ per cent over the H$\alpha$-emitting disc of
IC\,10. 

On the other hand, the foreground Galactic reddening $E(B-V)$ in
certain directions towards IC\,10 can be as low as 0.37 or as
high as 0.87 \citep[see][]{riche01}.  Thus, due to our assumed uniform value of
$E(B-V) = 0.77$ towards IC\,10, there can a systematic error in the range $-60$
and $+25$ per cent to the thermal emission.

The unknown $T_{\rm e}$ can give rise to up to $\sim 10$ and $20$ per cent
error in the estimated thermal emission at 0.32 and 6.2 GHz, respectively
\citep[see][for details]{tabat07b}. Overall, the estimated thermal emission and
thereby $\fth$ can have a systematic error up to $\sim30$ per cent at 0.32 GHz
and $\sim50$ per cent at 6.2 GHz. However, errors in the thermal fraction
affect the non-thermal emission less severely. For example, within the
H$\alpha$-emitting disc of IC\,10, an $f_{\rm th, 6.2 GHz}~(f_{\rm th, 0.32
GHz})$ of 0.5 (0.2) with an error of $30$ per cent gives rise to a 30 (10) per
cent error on the non-thermal emission. 

Clearly, a large error in a high $\fth$ region will affect the non-thermal
emission most. In Section~\ref{abs_ff} though, we show that the H{\sc ii}
regions showing 100 per cent thermal emission agree with our estimated thermal
flux. Therefore, we believe that our estimated non-thermal emission maps do not
suffer from large systematic errors.

\begin{figure*}
\begin{centering}
\begin{tabular}{c}
{\mbox{\includegraphics[width=12cm, trim=0mm 0mm 110mm 5mm,clip]{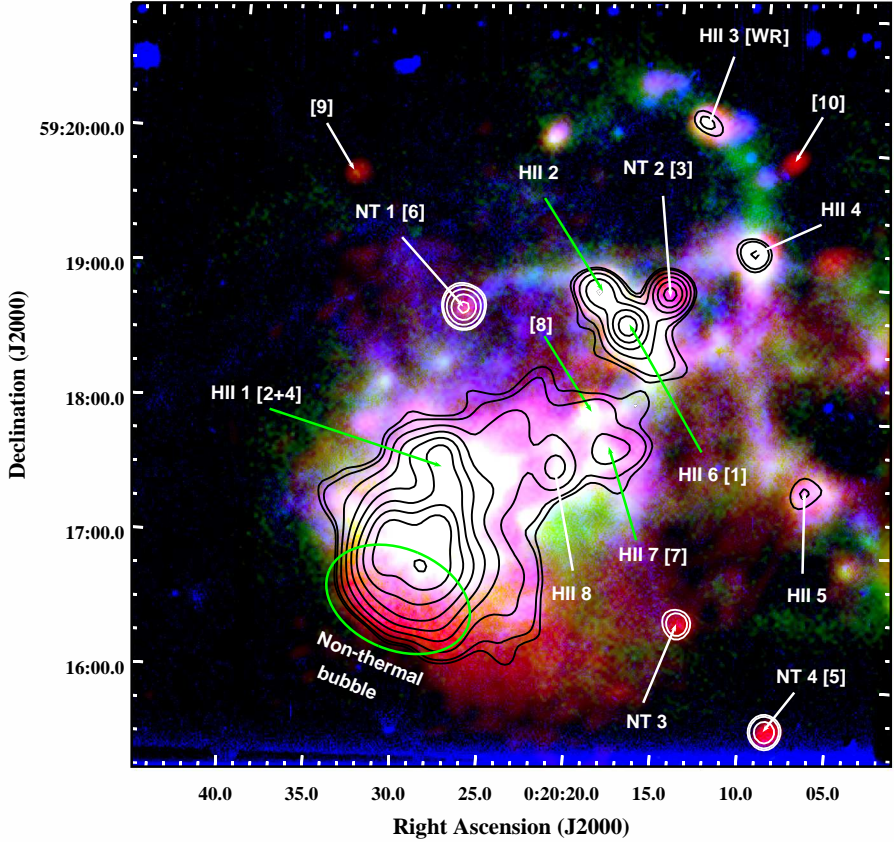}}} \\
\end{tabular}
\end{centering}
\caption{Various point sources detected at 0.32 GHz are marked on the composite
image of IC\,10 where the H$\alpha$, $70\,\mu$m and 6.2 GHz emissions are shown
in blue, green and red colours, respectively. The contour levels show the total
intensity at 0.32 GHz at $(8, 9, 13, 19, 27.5, 38.5, 52, 68, 86, 107)\times
150~\mu$Jy beam$^{-1}$. The sources marked as H{\sc ii} are the H{\sc ii}
regions, while the sources marked as NT are non-thermal emitting regions
without any H$\alpha$ counterpart. The alternate nomenclature
corresponding for some of these sources detected in the e-Merlin observations
of \citet{westc17} are given within the square brackets.}
\label{composit}
\end{figure*}

\subsection{Star formation rate} \label{sfr}

To trace the recent star formation rate (SFR; $\lesssim10$ Myr), we used the
extinction corrected H$\alpha$ flux (in Eq.~\ref{eqn_hacorr}) to convert it
into SFR by following the calibration given in \citet{kenni12} based on
\citet{hao11}. It should be noted that the calibration is valid for star
formation in solar metallicity environments. In order to account for the low
metallicity of IC\,10, we multiply the H$\alpha$ flux by a factor 0.89 before
converting it into a SFR. This essentially accounts for the fact that, for a
given SFR in a low metallicity environment, the escape fraction of H$\alpha$
photons is larger as compared to a higher metallicity environment. The factor
0.89 was deduced from \citet{raite10} who estimated emergent fluxes in
sub-solar metallicity environments for a Salpeter initial mass function
\citep[IMF;][]{salpe55, chabr05} and constant star formation over at least the
last $10^8$ years. The galaxy-integrated SFR was estimated to be
$0.037\pm0.003$ $\rm M_\odot\, yr^{-1}$. We however note, although the
statistical error on our estimated SFR is low, the systematic error in the
conversion factor of H$\alpha$ flux to SFR can vary by up to a factor of
$\sim2$ \citep{weilb01, hao11}.

\section{Results}

\subsection{Total intensity}

The total intensity radio continuum map of IC\,10 at 0.32 GHz, shown in
Fig.~\ref{totI_ic10}, is the highest angular resolution and sensitivity map
available for IC\,10 at frequencies below 1 GHz to date. Overall, the radio
continuum emission of IC\,10 at 0.32 GHz originates from a variety of
structures and sources. In Fig.~\ref{composit}, we mark the point-like sources
and the non-thermal bubble detected at 0.32 GHz on a composite colour image of
IC\,10 with H$\alpha$ emission from the Perkins 1.8-m telescope in blue,
$70\,\mu$m emission from {\it Herschel} PACS in green and 6.2 GHz emission from
VLA+Effelsberg in red. The H{\sc ii} regions, mostly seen as white regions in
the figure are marked as H{\sc ii} 1--8. The brighter radio continuum emission
regions in IC\,10 are coincident with high SFR regions and giant molecular
clouds (GMCs) detected in the survey of CO($J=1\to0$) emission by
\citet{leroy06}.  However, not all bright radio continuum emission originates
from star forming regions, such as the non-thermal sources appearing reddish in
Fig.~\ref{composit} that are marked as NT 1--4. Several of these compact
emitting regions are detected at 1.5 GHz in the high-resolution e-MERLIN
observations by \citet{westc17}. The alternate nomenclature of the e-MERLIN
detected sources are given within square brackets.

\begin{table}
 \centering
  \caption{Integrated flux density of IC\,10.}
   \begin{tabular}{@{}ccl@{}}
  \hline
 Frequency &  Flux density & References\\
  (GHz)    & (Jy)   &\\
 \hline
0.32  & $0.545\pm0.025$  & This paper (GMRT)\\
1.415 & $0.304\pm0.030$   & 1\\
1.42$^\dagger$  & $>0.270$   & This paper (VLA C-array)\\
1.43$^\P$  &  $0.377\pm0.011$  & 2\\
1.49  & $0.30\pm0.02$   & NVSS\\
2.64  & $0.250\pm0.02$   & 3 \\ 
4.85  & $0.222^{+0.015}_{-0.025}$  & 7\\ 
6.2$^\dagger$   &  $0.131\pm0.07$   & 6 \\
6.2   & $0.186\pm0.010$  & This paper (VLA+Effelsberg)\\
8.35  & $0.183\pm0.008$  & 2\\
10.45 & $0.155\pm0.016$  & 4\\
10.7  & $0.165\pm0.007$  & 5\\
24.5  & $0.118\pm0.018$  & 7\\
\hline
\end{tabular}

\begin{flushleft}
$^1$\cite{shost74},
$^2$\cite{chyzy16},
$^3$\cite{chyzy11},
$^4$\cite{chyzy03},
$^5$\cite{klein83},
$^6$\cite{heese11},
$^7$\cite{klein86}.\\
$^\dagger$ These points are not included while fitting the spectrum
as they suffer from missing flux density.\\
$^\P$ This point is not included while fitting the spectrum as this
includes contribution from the halo unlike other data.
\end{flushleft}
\label{tableflux}
\end{table}

\begin{figure*}
\begin{centering}
\begin{tabular}{cc}
{\mbox{\includegraphics[width=8cm,trim=10mm 10mm 0mm 5mm, clip]{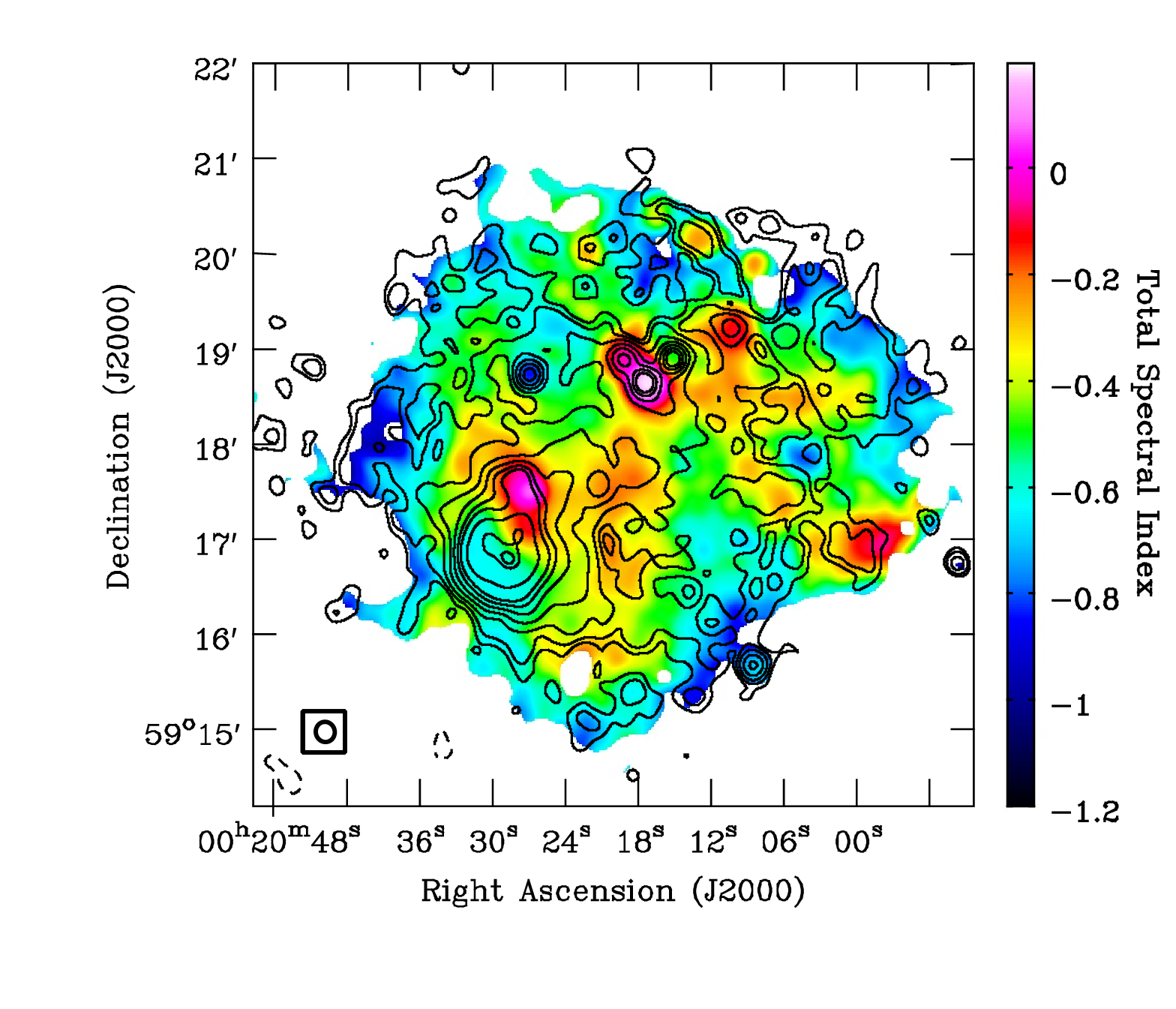}}}&
{\mbox{\includegraphics[width=8cm,trim=10mm 10mm 0mm 5mm, clip]{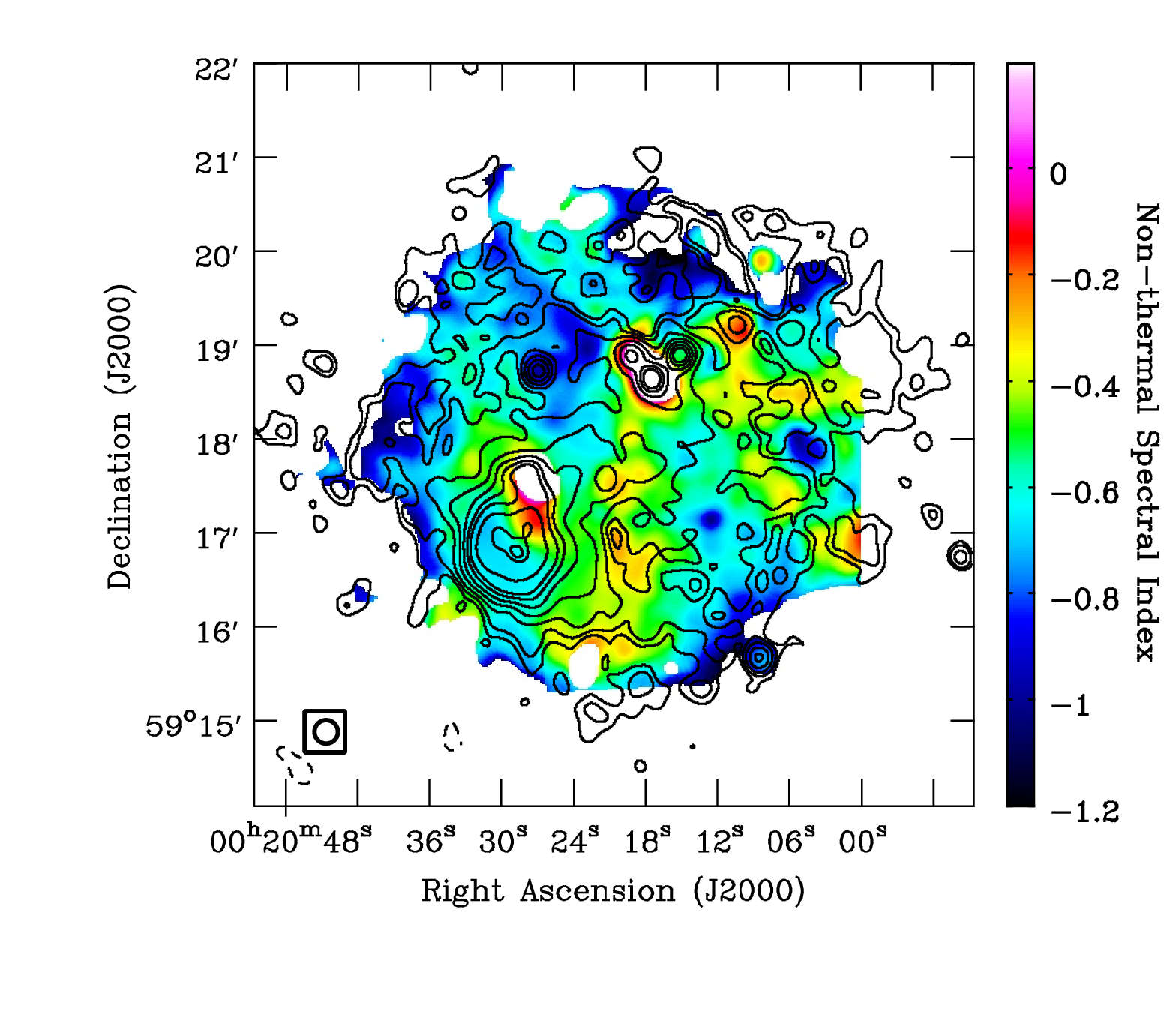}}}\\
\end{tabular}
\end{centering}
\caption{{\it Left:} Total spectral index map of IC\,10 computed between 0.32
and 6.2 GHz using the total radio continuum emission. Both the maps were convolved
to a common resolution of 15 arcsec and only the pixels $>3\sigma$ were
considered. {\it Right:} Non-thermal spectral index map ($\ant$) between 0.32
and 6.2 GHz at 15 arcsec angular resolution. The inner white regions are
blanked where the total radio continuum emission is entirely thermal in nature.
Pixels with $\ant$ in the range $-0.4$ and $-0.1$ likely occur due to improper
separation of the thermal emission. The contours are the total intensity
contours at 0.32 GHz, same as Fig.~\ref{totI_ic10}.}
\label{spind}
\end{figure*}

Among the detailed studies at higher frequencies mentioned in
Section~\ref{studies}, at 1.4 GHz \citet{chyzy16} found evidence of a
synchrotron halo extending up to $\sim2.5$ kpc likely originating from
magnetized winds driven by star formation. The rms noise of the map was 28
$\mu$Jy beam$^{-1}$ at an angular resolution of 26 arcsec. We made a lower
resolution image of IC\,10 at 0.32 GHz using baselines shorter than $\rm
10~k\lambda$ and convolved to 26 arcsec which rendered an image rms noise of
$540~\mu$Jy beam$^{-1}$. At our sensitivity, the synchrotron halo remains
undetected at 0.32 GHz. This implies that the spectral index of the halo around
the $8\sigma$ contour of \citet{chyzy16} map is $\gtrsim-1.0$ between 0.32 and
1.4 GHz for it to remain undetected at 0.32 GHz assuming a $2\sigma$ threshold.
Thus, the CREs giving rise to the radio continuum emission are unlikely to be
significantly affected by synchrotron and/or inverse-Compton losses.
\citet{chyzy16} suggested that a galactic wind with velocities $\sim25~\rm
km\,s^{-1}$ is responsible for the extended radio halo, i.e., the CREs would
require $\sim60$ Myr to travel to the edge of the halo at $\sim1.5$ kpc (a wind
velocity of $\sim60~\rm km\,s^{-1}$ in turn would only need $\sim25$ Myr to
reach the same distance). Thus, the magnetic field strengths in the halo must
be $\lesssim 5~\mu$G for the transport time-scales to be less than that of the
synchrotron loss time-scale for the CREs emitting at 1.4 GHz to remain
unaffected by synchrotron losses.

At 6.2 GHz, \citet{heese11} found the radio continuum emission to closely trace
the star forming regions.  In particular, the H{\sc ii} regions visible in the
H$\alpha$ images are also bright at 6.2 GHz (marked as H{\sc ii} $1-8$ in
Fig.~\ref{composit}). They also found their radio continuum spectrum to be
consistent with thermal emission, with $\alpha$ in the range $-0.1$ to $-0.2$.
At 0.32 GHz, we detect these bright star-forming regions as well. Apart from
these H{\sc ii} regions, we also detect bright point-like sources in IC\,10
that do not have any H$\alpha$ counterpart and are likely to be non-thermal
sources (marked as NT $1-4$ in Fig.~\ref{composit}).

\subsection{Thermal/non-thermal emission and spectral index}

In Fig.~\ref{integflux}, we plot the galaxy-integrated flux density of IC\,10
between 0.32 and 24.5 GHz based on our measurements and data collected from the
literature listed in Table~\ref{tableflux}. Overall, the radio continuum
spectra of IC\,10 is found to be flatter with spectral index $-0.34\pm0.01$
than the $-0.7$ or steeper found in normal spirals. The green dotted line shows
the estimated thermal emission as described in Section~\ref{thermalsep}. The
estimated $\fth$ of $0.2\pm0.05$ at 0.32 GHz is significantly higher than what
is observed in normal star-forming galaxies at these frequencies \citep{basu12}
and this high contribution of thermal free--free emission flattens the
integrated spectra. The red points are the non-thermal flux density after
subtracting the thermal emission at each frequency. The red dashed line is the
best fit power-law to the non-thermal flux densities in the $\log$--$\log$
space and has a non-thermal spectral index\footnote{We distinguish between
total and non-thermal spectral index as $\alpha$ and $\ant$, respectively.}
($\ant$) of $-0.55\pm0.04$. In order to assess any curvature in the non-thermal
spectrum originating due to synchrotron and/or inverse-Compton losses, we also
fitted a second-order polynomial of the form $\log S_{\rm \nu, nt} = s_0 + \ant
\log \nu + \beta\, (\log \nu)^2$. The non-thermal radio continuum spectrum
shows an indication of curvature and is shown as the blue dot-dashed curve in
Fig.~\ref{integflux}. We note that a curved spectrum fits the data marginally
better than a single power-law. From the fit, the injection spectral index is
found to be $-0.45\pm0.05$ and is consistent with that of a fresh CRE population
generated by diffusive shock acceleration \citep{bell78I, bland87}.

\begin{figure}
\begin{centering}
\begin{tabular}{c}
{\mbox{\includegraphics[width=9cm, trim=5mm 5mm 0mm 0mm,clip]{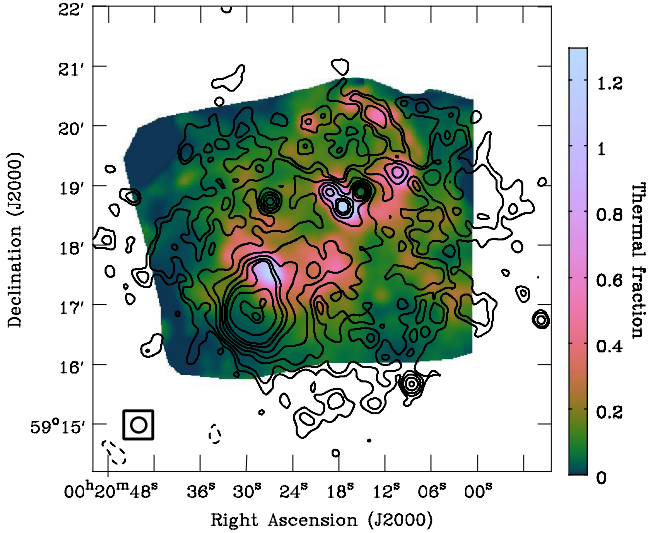}}} \\
\end{tabular}
\end{centering}
\caption{Thermal fraction map estimated at 0.32 GHz at 15 arcsec resolution.
Overlaid contours are the total intensity 0.32 GHz emission same as
Fig.~\ref{totI_ic10}.} 
\label{fth_325}
\end{figure}

In Fig.~\ref{spind}, left- and right-hand panels show the total and
non-thermal spectral index maps of IC\,10, respectively, estimated between 0.32
and 6.2 GHz. The bright point sources detected at 0.32 GHz, coincident with
H$\alpha$ emission (marked as H{\sc ii} $1-8$ in Fig.~\ref{composit}), have
$\alpha$ values close to $-0.1$ or higher. This indicates that such sources are
direct tracers of H{\sc ii} regions and are dominated by thermal free--free
emission. In fact, the sources marked as H{\sc ii} 1, 2 and 6 exhibit
broad-band spectra consistent with 100 per cent thermal emission. In
Fig.~\ref{fth_325}, we show the map of the thermal fraction at 0.32 GHz ($f_{\rm
th, 0.32GHz}$) estimated by extrapolating the thermal emission with a constant
thermal spectral index of $-0.1$. Clearly, $f_{\rm th, 0.32 GHz}$ in the
location of H{\sc ii} 1, 2, 4 and 6, i.e., the whitish regions in
Fig.~\ref{fth_325} are close to or exceed unity. We discuss the nature of the
radio continuum spectra of these regions in detail in Section~\ref{abs_ff}. As
is also evident from the spectral index map, the regions with
$\alpha\gtrsim-0.3$ in the disc of IC\,10 closely follow the H$\alpha$
emission. The flat spectra are due to significant thermal emission where we
find $f_{\rm th, 0.32GHz}$ in the range 0.2 to 0.5 (see Fig.~\ref{fth_325}).
The $\ant$ in these regions are higher with values in the range $-0.5$ to
$-0.6$, close to the injection spectral index of fresh CREs.  The non-thermal
radio emission from the H$\alpha$ emitting disc likely originates from young
CREs produced during the star-burst.

The regions in the outer parts, i.e., outside the H$\alpha$ emitting disc, show
a steeper radio continuum spectrum with $\alpha$ in the range $-0.4$ to $-0.8$.
The thermal fractions are comparatively lower in the outer parts with $\fth$
typically lying between 0.01 to 0.1 at 0.32 GHz and 0.1 to 0.2 at 6.2 GHz. The
non-thermal spectrum in these regions is steep with $\ant \lesssim -0.8$,
indicating the CREs are likely affected by synchrotron losses.

\begin{figure}
\begin{centering}
\begin{tabular}{c}
{\mbox{\includegraphics[width=8.5cm, trim=8mm 0mm 13mm 10mm, clip]{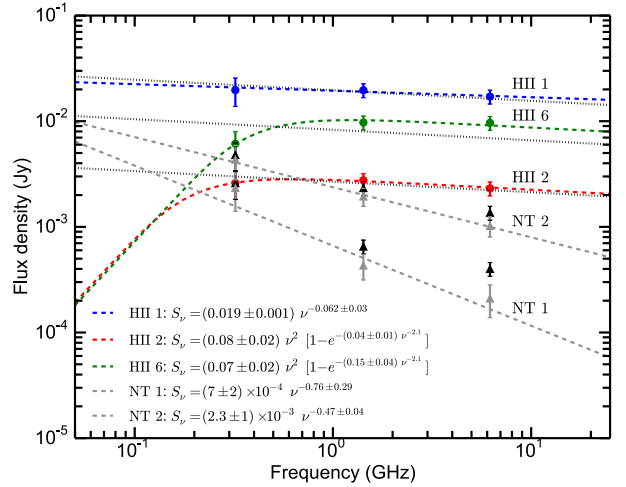}}}
\end{tabular}
\end{centering}
\caption{Total intensity spectrum of some of the bright compact sources
detected at 0.32 GHz marked in Fig.~\ref{composit}. The circles represents the
sources H{\sc ii} 1, 2 and 6 and are coincident with strong H$\alpha$ emission.
The black triangles represents the observed total flux densities of the sources NT
1 and 2 which do not show enhanced H$\alpha$ emission. The grey triangles shows
the non-thermal flux density after subtracting the thermal emission estimated
at the positions of NT 1 and 2. All the errors are shown at $3\sigma$
level. The various dashed lines are the best representation of the spectrum
for the corresponding sources. The dotted lines for the sources H{\sc ii} 1, 2
and 6 are the extrapolated thermal flux density estimated using H$\alpha$
emission.}
\label{totI_hII}
\end{figure}

\subsubsection{Free--free absorption} \label{abs_ff}

Some of the bright H$\alpha$ emitting regions show beam-averaged $f_{\rm
th, 0.32 GHz}$ in the range 0.9 and 1.2, while at 6.2 GHz, no such region shows
$f_{\rm th,6.2GHz}>1$. $\alpha$ in these regions have values $-0.1$ or higher.
A close inspection of the total radio continuum emission at 0.32, 1.43 and 6.2
GHz reveals that the regions H{\sc ii} 2 and 6 shows the effects of thermal
free--free absorption which dominates at lower radio frequencies. In
Fig.~\ref{totI_hII}, we show the spectrum based on measurements at three
frequencies of the brightest\footnote{The flux densities of the fainter
point-like sources identified in Fig.~\ref{composit} are confused with the
background diffuse emission and are unsuitable for a detailed study.} H{\sc ii}
regions, H{\sc ii} 1, 2 and 6 and non-thermal sources NT 1 and 2. The sources
H{\sc ii} 2 and 6 clearly show lower radio continuum flux densities at 0.32 GHz
compared to the extrapolated thermal emission estimated from the H$\alpha$
emission (shown as the dotted lines in Fig.~\ref{totI_hII}), thereby
giving rise to $f_{\rm th, 0.32 GHz}>1$ and $\alpha > -0.1$.  For these two
sources, the spectrum is best represented by an optically thick thermal
free--free spectrum of the form $S_\nu = A\,\nu^2\,(1-e^{-m\,\nu^{-2.1}})$.
While for the source H{\sc ii} 1, the spectrum is consistent with an optically
thin thermal free--free spectrum of the form $A\,\nu^{-0.1}$. The best-fit
values of the parameters $A$ and $m$ are shown in Fig.~\ref{totI_hII}. 

For the sources H{\sc ii} 2 and 6, comparing the best-fit parameter $m$ to
Eqs.~\ref{eqn_sth} and \ref{tauff}, we estimate the thermal free--free emission
to be optically thick ($\tau_{\rm ff} \geq 1$) below $0.22\pm0.03$ and
$0.41\pm0.05$ GHz, respectively. Recently, \citet{hinds16} in a study of
Galactic H{\sc ii} regions, found evidence of a turnover in the free--free
spectrum typically with turnover frequencies in the range 0.3 to 1.2 GHz.  We
estimated\footnote{The parameter $m$ is related to the EM as follows: $m =
0.082\, T_{\rm e}^{-1.35}\, \mathrm{EM}$ (cf. Eq.~\ref{tauff}).} the EM to be
$(1.2 \pm 0.3) \times 10^5$ and $(4.6 \pm 1.2) \times 10^5~\rm pc\,cm^{-6}$ for
H{\sc ii} 2 and 6, respectively which in turn is related to the average
thermal electron density $\langle n_{\rm e}\rangle$ as \citep{berkh06}:
\begin{equation}
\langle n_{\rm e} \rangle = \left(\frac{{\rm EM}\,{f}}{h_{\rm HII}}\right)^{1/2},
\end{equation}
where, $f$ is the filling factor and $h_{\rm HII}$ is the size of the H{\sc ii}
region along the line of sight. Assuming a typical $f \sim 5$ per cent for the
clumpy star-forming disc \citep{ehle93} and $h_{\rm HII} \sim 50$ pc, i.e., the
spatial resolution of our observations, we estimate $\langle n_{\rm e} \rangle$
to be $\sim 11$ and $22~\rm cm^{-3}$ for H{\sc ii} 2 and 6, respectively, which
are typical values for H{\sc ii} regions \citep{hunt09}. \citet{deeg93} pointed
out that such values of EM and $n_{\rm e}$ are necessary to explain the radio
continuum spectra of star-forming blue compact dwarf galaxies. To better
constrain $\langle n_{\rm e} \rangle$ and the sizes of the H{\sc ii} regions,
high resolution radio continuum observations at even lower frequencies with 
LOFAR will be necessary.

The thermal emission independently estimated using extinction corrected
H$\alpha$ emission well represents the emission in the three H{\sc ii}
regions (shown as black dotted lines in Fig.~\ref{totI_hII}). This suggests
that the extinction corrected H$\alpha$ emission is a good tracer of the
thermal free--free emission on scales below 100 pc.  Since, the total radio
continuum emission in these regions are already consistent with 100 per cent
thermal emission, it is difficult to estimate the contribution of the
non-thermal component. We have therefore blanked these pixels in our future
calculations.

\begin{figure}
\begin{centering}
\begin{tabular}{c}
{\mbox{\includegraphics[width=8cm,trim=0mm 10mm 10mm 5mm]{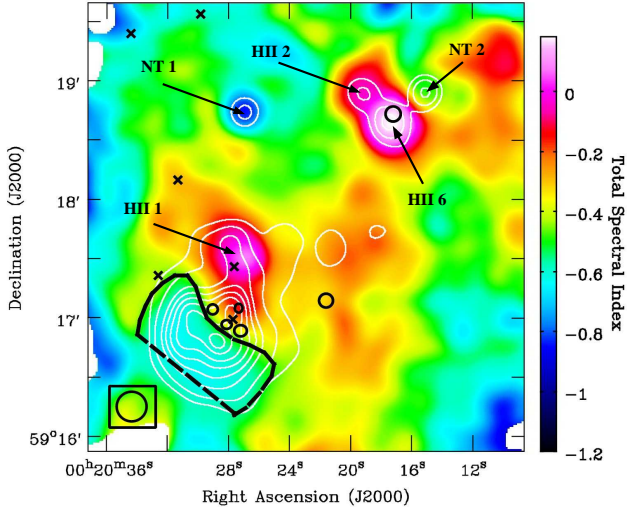}}}
\end{tabular}
\end{centering}
\caption{Total spectral index map between 0.32 and 6.2 GHz of the inner regions
of IC\,10. The locations of the GMCs found in the survey by Leroy et al. (2006)
are shown as circles and crosses. The size of the circles are the same as the
physical sizes estimated by Leroy et al. (2006), while for the crosses the
size information was not available. The black line shows the sharp edge
observed in the spectral index map coincident with the boundary of the
molecular clouds. We also mark the compact sources marked in
Fig.~\ref{composit}. The contours are the total intensity contours at 0.32 GHz
drawn at $(15, 25, 40, 50, 60, 70, 80, 85)\times 150~\mu$Jy beam$^{-1}$.}
\label{spind_zoom}
\end{figure}

\subsubsection{Non-thermal bubble}

An interesting feature observed in the maps of $\alpha$ and $\ant$ lies towards
the south-eastern edge of IC\,10 marked as the ``non-thermal bubble'' in
Fig.~\ref{composit}. The bubble was first identified in the radio continuum
observations of \citet{yang93}. The sharp boundary of this region is
distinctly visible only in the spectral index maps (see Fig.~\ref{spind}). The
total intensity emission is merged with the bright H{\sc ii} region H{\sc ii} 1
and does not show the bubble distinctly. In Fig.~\ref{spind_zoom} we show a
zoomed-in view of the spectral index map. We mark the sharp boundary of the
bubble and determine its center to be at RA=$00^{\rm h} 20^{\rm m} 29^{\rm s}$
and Dec.=$+59^{\circ} 16^{\prime} 39.5^{\prime\prime}$. The non-thermal bubble
extends $\sim300$ pc in the north-east to south-west direction and $\sim100$ pc
in the north-west to south-east direction. Its north-western edge, where the
spectrum flattens from being non-thermal with $\alpha\sim-0.6$ to thermal
$\alpha \sim -0.1$, is bound by giant molecular clouds (GMCs) as observed by
\citet{leroy06}, shown as the black circles in Fig.~\ref{spind_zoom}. This
suggests that the GMCs likely confine the CREs in the bubble.  This sharp
feature is also visible in the spectral index map computed between 1.43 and
4.86 GHz using a completely independent dataset \citep{chyzy16}.

The strongest non-thermal emission in IC\,10 originates from the bubble both at
0.32 and 6.2 GHz. $\fth$ in this region is lower than the other parts in the
disc of IC\,10 with $\fth \sim0.035$ and $0.16$ at 0.32 and 6.2 GHz,
respectively. Apart from the outer edges of IC\,10 where $\fth$ tends to be on
the low side, this is the only region within the H$\alpha$ emitting disc which
is dominated by non-thermal emission. The average $\alpha$ and $\ant$ between
0.3 and 6.2 GHz in the bubble is found to be $-0.61\pm0.02$ and $-0.66\pm0.03$,
respectively. Compared to the diffuse emission in the outer parts of IC\,10,
the non-thermal spectrum is slightly flatter in the bubble. Using broadband
data, \citet{heese15} modelled the radio continuum spectrum of the bubble using
a Jaffe--Perola model \citep{jaffe73}. Their modelling yielded an injection
spectral index of $-0.6\pm0.1$, consistent with the non-thermal spectral index
measured in our observations. This indicates that the CREs giving rise to the
synchrotron emission in the bubble are freshly generated.

\subsubsection{Non-thermal point-like sources}

Apart from the bright H{\sc ii} regions, we also find a few bright knots at
0.32 GHz which do not show strong H$\alpha$ emission. Unlike the diffuse
non-thermal bubble, these sources are compact and appear to be point-like
(marked as NT 1, 2, 3 and 4 in Fig.~\ref{composit}). These regions show locally
enhanced non-thermal emission and the thermal fractions are low with $f_{\rm
th, 0.32 GHz}$ in the range 0.05--0.08 (see Fig.~\ref{fth_325}). The
spectral index of their total radio continuum emission lies in the range $-0.8$
to $-0.5$, which is smaller than the main H$\alpha$ emitting disc of IC\,10.
The total intensity spectra for two of the sources NT 1 and 2 are shown in
Fig.~\ref{totI_hII} which are distinctly different from a thermal spectrum.
Because of their steep spectra, these point-like sources are not prominently
visible in the images at 1.42 and 6.2 GHz \citep{chyzy16, heese11}. Non-thermal
sources NT 1, 2 and 4 are detected by e-MERLIN observations and are shown to be
compact sources. They are likely to be background star-forming galaxies,
hosting active galactic nuclei \citep{westc17}.

\subsection{Magnetic field strengths}

\begin{figure}
\begin{centering}
\begin{tabular}{c}
{\mbox{\includegraphics[width=8cm,trim=18mm 12mm 0mm 5mm, clip]{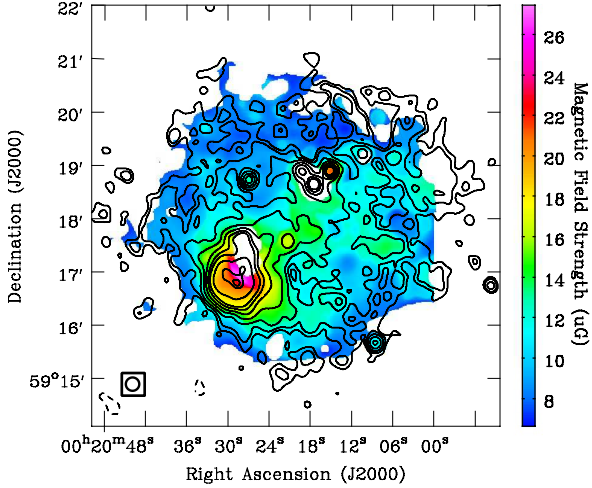}}}
\end{tabular}
\end{centering}
\caption{Total magnetic field strength of IC\,10 estimated assuming energy
equipartition between cosmic ray particles and magnetic field. The inner white
regions are blanked because of uncertainty in the estimation of non-thermal
emission. The overlaid contours are the total intensity emission at 0.32 GHz
(same as Fig.~\ref{totI_ic10}; see text for details).}
\label{bfield}
\end{figure}

We computed the total magnetic field strength ($B_{\rm tot}$) of IC\,10 on a
pixel-by-pixel basis using the non-thermal emission and $\ant$ maps assuming
energy equipartition between cosmic ray particles and the magnetic field
\citep{beckkrause05}. We assumed the ratio of number densities of relativistic
protons to electrons ($K$) to be 100 and the path-length of the synchrotron
emitting media ($l$) to be 1 kpc and corrected this for the galaxy's
inclination. In Fig.~\ref{bfield} we present the total magnetic field strength
map of IC\,10. The average field strength within the $3\sigma$ region (excluding
the background non-thermal sources, NT 1, 2 and 4 and the H{\sc ii} regions
with 100 per cent thermal emission) is estimated to be $\sim10~\mu$G. Our
estimated field strengths are lower than the average field strength of
$14~\mu$G estimated by \citet{chyzy16}. Note that, \citet{chyzy16} found the
field to be strongest ($\sim29~\mu$G) in the H{\sc ii} complex (marked as H{\sc
ii} 1 in Fig.~\ref{composit}). However, our study shows that this region has
100 per cent thermal emission and the assumption of energy equipartition is
invalid and hence has been blanked in Fig.~\ref{bfield}. 

We find the magnetic fields to be the strongest in the non-thermal bubble with
field strengths of $\sim20~\mu$G, similar to \citet{chyzy16}. However, the
field strengths could be higher in this region as the bubble is unlikely to
extend 1 kpc along the line of sight \citep[see][]{heese15}. 

In the H$\alpha$-emitting disc, the field strength lies in the range 10 to
$15~\mu$G and falls off to $\sim8~\mu$G in the outer parts. Excluding the
non-thermal bubble, the average field is estimated to be $\sim12~\mu$G in the
mid-plane. The estimated magnetic field strengths in IC\,10 are larger
than those typically observed in dwarf irregular galaxies ($\lesssim 5~\mu$G;
\citealt{chyzy11}) and comparable to those observed in normal spiral galaxies
($9-15~\mu$G; \citealt{basu13}).

Note that, our estimated magnetic field strengths can be scaled by a
factor $[10^{-2} (K+1)\times \Delta f_{\rm nth}/l]^{1/(3 - \ant)}$ because of
our assumed values of $K$ and $l$. Thus, a factor of 2 difference in the path
length would give rise to a maximum of $\sim 22$ per cent systematic error in
the magnetic field strength. On the other hand, the statistical error of the
magnetic field strength depends on the signal-to-noise ratio (SNR) of the
non-thermal emission. Typically, in the high SNR regions ($\gtrsim 5$), i.e.,
within the H$\alpha$-emitting disc of IC\,10, the error lies in the range
$2-10$ per cent, while in the outer parts with SNR 3--5, the error can be up to
$\sim20$ per cent.\footnote{The errors were computed using the Monte Carlo method
described in \citet{basu13}.} Further, to include an error in the estimated
thermal emission, we consider the term $\Delta f_{\rm nth}$ defined as $\Delta
f_{\rm nth} = 1 \pm d f_{\rm nth}/f_{\rm nth}$, where $d f_{\rm nth}$ is the
error on the estimated non-thermal fraction ($f_{\rm nth} = 1 - \fth$). In
terms of $\fth$, $\Delta f_{\rm nth}$ is given as $\Delta f_{\rm nth} = [1 - (1
\pm a)\fth]/[1 - \fth]$, where $a$ is the relative error on $\fth$. An error
of 30 per cent on $\fth$ (i.e., $a = \pm0.3$) in a region with $\fth = 0.5
(0.2)$ gives rise to less than $10 (3)$ per cent error on the magnetic field
strength. Overall, the error on the estimated magnetic field strength is
$\lesssim 10$ per cent within the H$\alpha$-emitting disc of IC\,10 and up to
$\sim 20$ per cent in the outer parts.

\begin{figure}
\begin{centering}
\begin{tabular}{c}
{\mbox{\includegraphics[width=8cm, trim=10mm 0mm 0mm 0mm,clip]{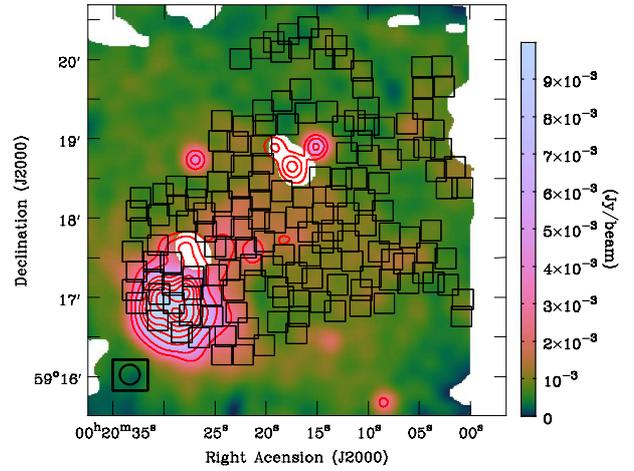}}} \\
\end{tabular}
\end{centering}
\caption{Non-thermal emission map at 0.32 GHz at 15 arcsec resolution. The
inner white pixels are blanked as they correspond to the location of H{\sc ii}
regions with 100 per cent thermal emission (see Section~\ref{abs_ff}). The
boxes represent regions within which the our spatially resolved studies are
performed and are roughly of one beam-size. Overlaid red contours are the total
intensity 0.32 GHz emission same as Fig.~\ref{spind_zoom}.}
\label{regions}
\end{figure}

\begin{figure*}
\begin{centering}
\begin{tabular}{cc}
{\mbox{\includegraphics[width=7cm,trim=0mm 0mm 10mm 35mm, clip]{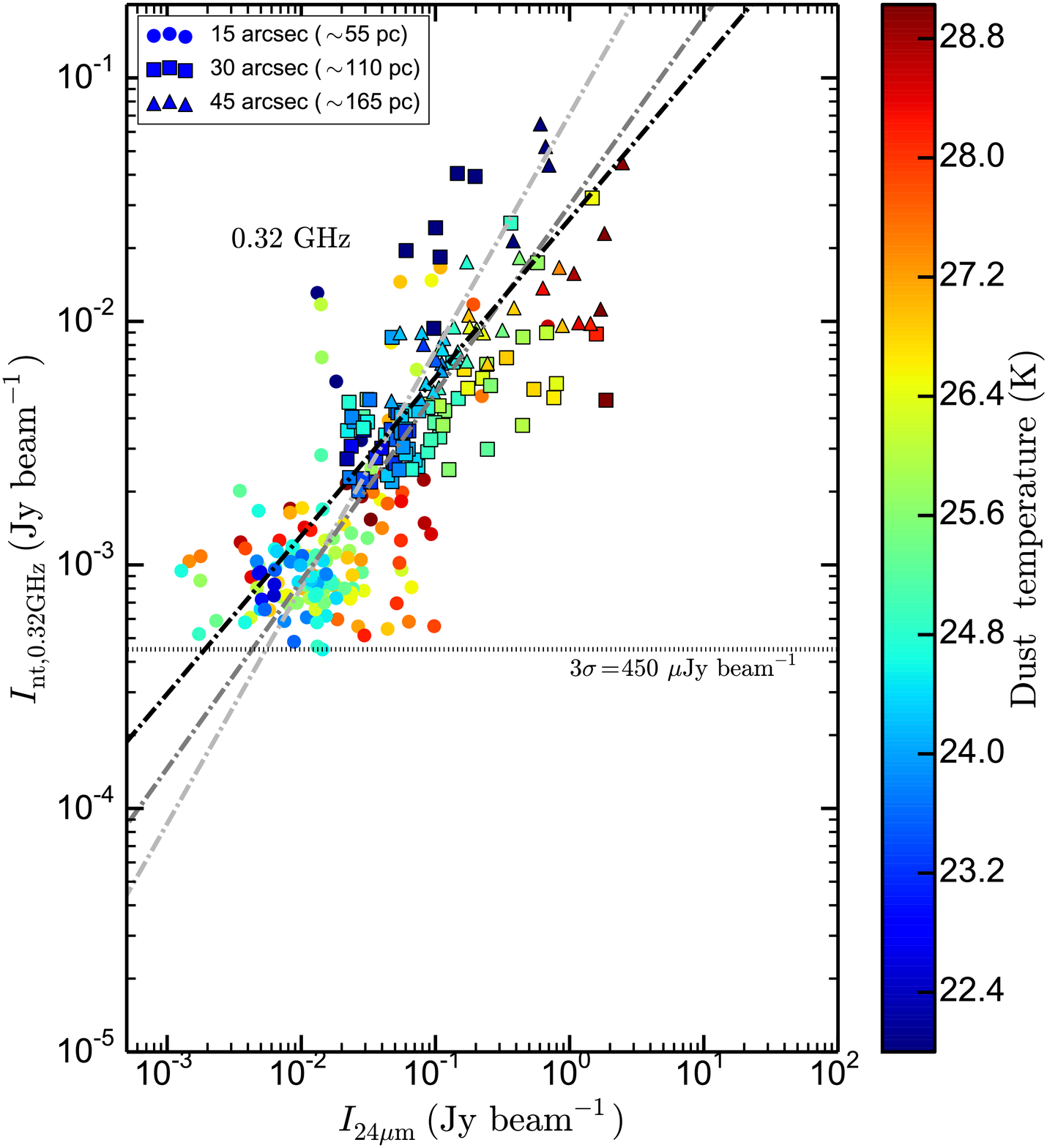}}}&
{\mbox{\includegraphics[width=7cm,trim=0mm 0mm 10mm 35mm, clip]{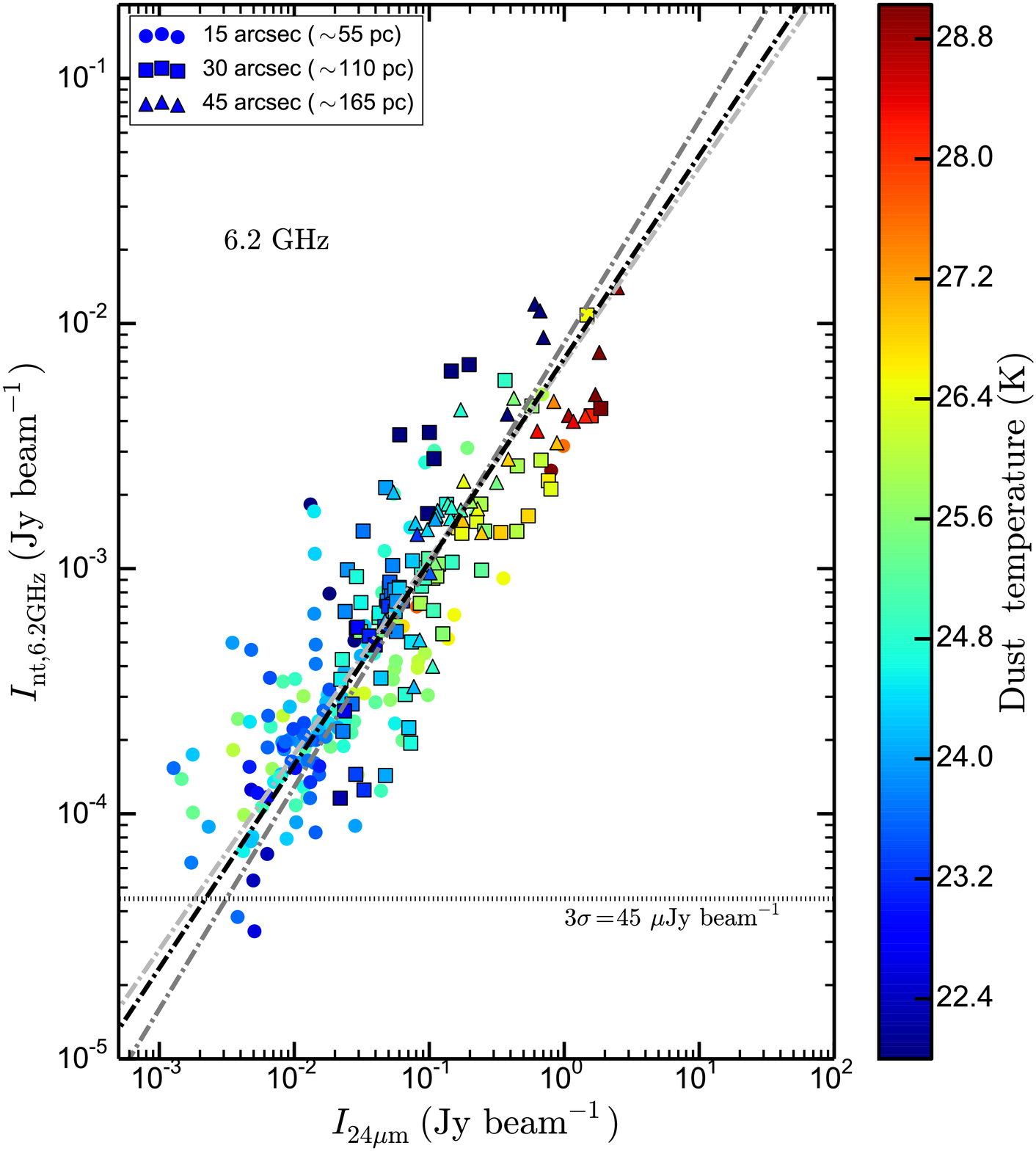}}}\\
{\mbox{\includegraphics[width=7cm,trim=0mm 0mm 10mm 35mm, clip]{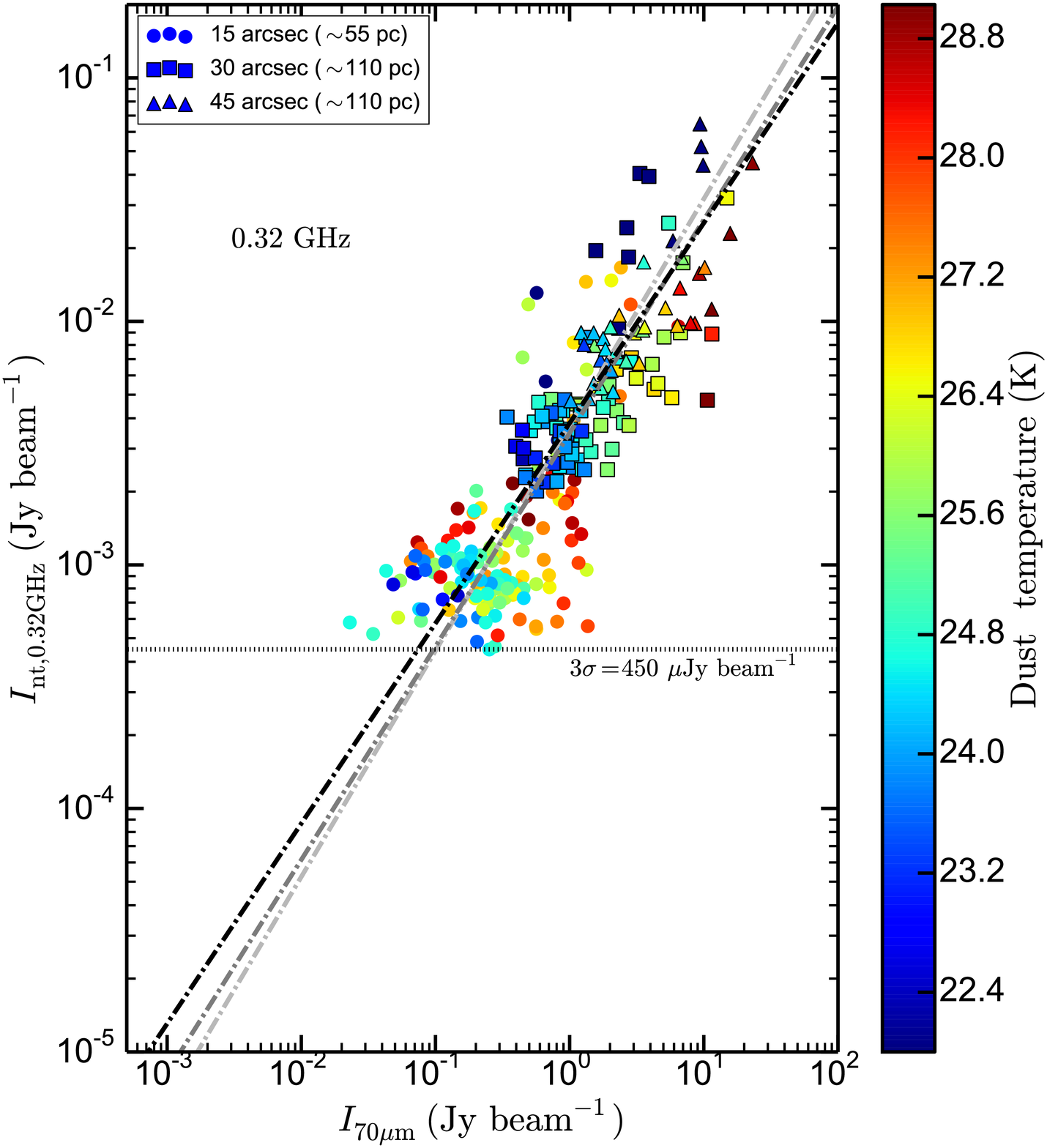}}}&
{\mbox{\includegraphics[width=7cm,trim=0mm 0mm 10mm 35mm, clip]{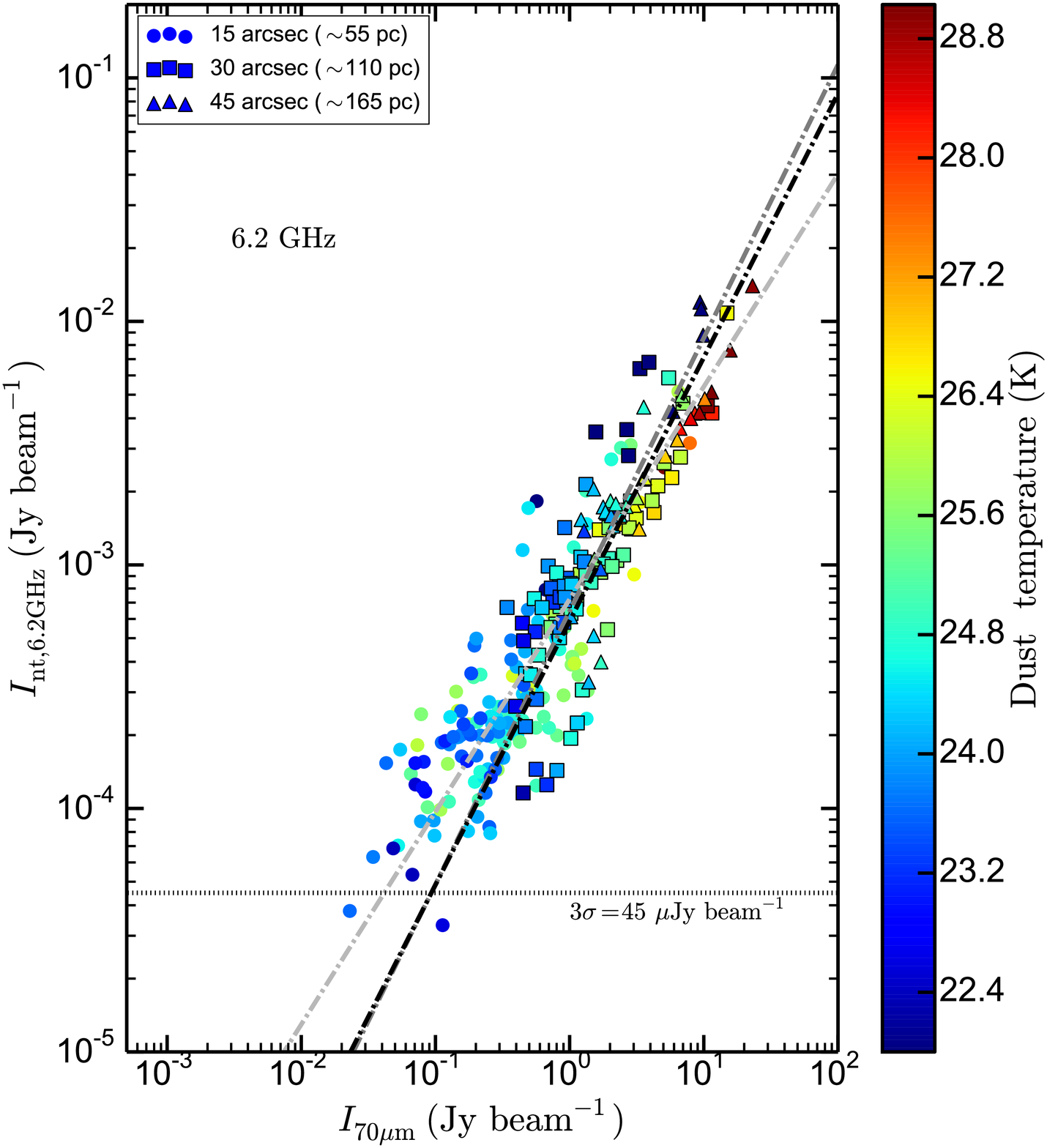}}}\\
\end{tabular}
\end{centering}
\caption{Spatially resolved non-thermal radio continuum intensity vs.  MIR
intensity at $24\,\mu$m (top panels) and, vs. FIR
intensity at $70\,\mu$m (bottom panels). The left-hand and right-hand sides are
for the emission at 0.32 and 6.2 GHz, respectively. The circle, square and
triangle symbols are averaged over 15, 30 and 45 arcsec apertures,
respectively. The dash-dotted lines are the best fit lines in the $\log-\log$
space, where the light grey, grey and black lines are for 15, 30 and 45 arcsec,
respectively. The $3\sigma$ detection threshold for the non-thermal emission
is shown as a dotted line. The points are coloured based on
the dust temperature.}
\label{radioFIR}
\end{figure*}

\begin{table*} \centering 
 \caption{Summary of spatially resolved radio--infrared relation in IC\,10.
The slopes ($b$) and $\log a$ are obtained by fitting the non-thermal radio
continuum {\it versus} IR intensities as $I_{\rm nt, \nu} = a \times I_{\rm
IR}^b$ in the $\log$--$\log$ space. $\sigma_{\rm IR}$ is the
dispersion around the fit after normalizing $x-$ and $y-$axis to their median
values and is a measure of the scatter or ``tightness'' of the corresponding
relation.}
\begin{tabular}{@{}ccccccccccccccc@{}} 
 \hline 
  \multicolumn{1}{c}{}&
  \multicolumn{14}{c}{Spatial scale}\\
  \cline{2-15}
  \multicolumn{1}{c}{}&
  \multicolumn{4}{c}{55 pc} &
  \multicolumn{1}{c}{}&
  \multicolumn{4}{c}{110 pc} &
  \multicolumn{1}{c}{}&
  \multicolumn{4}{c}{165 pc} \\
  \cline{2-5}
  \cline{7-10}
  \cline{12-15}
 Frequency   &  $r_S$    & Slope & $-\log\,a$ & $\sigma_{\rm IR}$ & & $r_S$  & Slope & $-\log\,a$ & $\sigma_{\rm IR}$ & & $r_S$ & Slope &  $-\log\,a$ & $\sigma_{\rm IR}$\\
 \hline

  \multicolumn{1}{c}{}&
  \multicolumn{14}{c}{\large Using 24 $\mu$m MIR emission}\\
0.32 GHz    & 0.47   & $0.97\pm0.15$ & $1.15\pm0.28$ & {4.36}  & & 0.52  & $0.77\pm0.13$ & $1.52\pm0.16$ & {2.51} & & 0.68 & $0.65\pm0.10$ & $1.58\pm0.09$ & {1.34} \\ 
6.2 GHz    & 0.73   & $0.80\pm0.06$ & $2.16\pm0.11$ & {2.63} & & 0.73  & $0.91\pm0.09$ & $2.08\pm0.11$ & {2.38} & & 0.82 & $0.83\pm0.10$ & $2.14\pm0.07$ & {1.34} \\ 
    &    &  &  &  & &   &  &  &  & &  &  &  & \\ 
  \multicolumn{1}{c}{}&
  \multicolumn{14}{c}{\large Using 70 $\mu$m FIR emission}\\
0.32 GHz    & 0.62   & $0.93\pm0.09$ & $2.43\pm0.06$ & {2.42}  & & 0.68  & $0.88\pm0.11$ & $2.46\pm0.02$ & {1.79} & & 0.79 & $0.82\pm0.11$ & $2.42\pm0.05$ & {1.08} \\ 
6.2 GHz    &  0.84   & $0.87\pm0.04$ & $3.14\pm0.03$ & {1.66} & & 0.83 & $1.12\pm0.09$ & $3.19\pm0.03$ &   {1.52} & & 0.89 & $1.08\pm0.11$ & $3.23\pm0.07$ & {0.93} \\ 
\hline 
\end{tabular}
\label{corrtable} 
\end{table*}

\subsection{Spatially resolved radio--infrared relations}

We studied the well known radio--infrared relations in IC\,10 at angular
resolutions of 15, 30 and 45 arcsec, corresponding to spatial scales of
$\sim55$, 110 and 165 pc, respectively. We used the non-thermal emission maps
at 0.32 and 6.2 GHz and mid- and far-infrared (MIR and FIR) maps at 24 and
$70\,\mu$m, respectively. The radio--FIR and radio--MIR relations are thought
to be of different physical origin. The radio--FIR relation arises primarily
due to the coupling between magnetic field and gas density \citep{nikla97b,
schle13, schle16}, while the radio--MIR relation is a consequence of star
formation \citep{heese14}. We therefore expect different dispersions for the
two types of relations.

The infrared maps were convolved to the resolution of the non-thermal radio
maps, i.e., 15 arcsec, using the convolution kernels given by \citet{anian11}.
Both the non-thermal radio and IR intensities were computed averaged over
regions of one beam size. To ensure independence, the regions were separated
roughly by one beam and only the pixels above $3\sigma$ rms noise of the
non-thermal radio maps were considered. The regions within which the
averaging was done are shown in Fig.~\ref{regions} and are overlaid on the
non-thermal emission map at 0.32 GHz at 15 arcsec resolution. To ensure that
our results are least affected due to insufficient thermal emission separation,
all the beams with $\ant\geq-0.45$ were excluded from our further analysis.
Further, we avoided all such beams which contained the non-thermal point-like
sources and H{\sc ii} regions with 100 per cent thermal emission. A similar
approach was followed for analysis at 30 and 45 arcsec angular resolutions.

In Fig.~\ref{radioFIR} we plot the non-thermal radio intensities versus the IR
intensities. The left- and right-hand panels are for 0.32 and 6.2 GHz,
respectively, while the top and bottom panels are {\it versus} 24 and
$70\,\mu$m, respectively. The circle, square and triangle symbols are averaged
over 15, 30 and 45 arcsec, respectively and they are coloured based on the dust
temperature ($T_{\rm dust}$; see Section~\ref{tight}). The error on the data
points are smaller than the scatter of the plots. The non-thermal emission at
6.2 GHz and the MIR emission at $24\,\mu$m of IC\,10 is found to be strongly
correlated with Spearman's rank correlation, $r_{\rm S}=0.73$ at $\sim55$ pc
spatial scale and increases slightly with $r_{\rm S} = 0.82$ at $\sim110$ and
165 pc scales. At 0.32 GHz, the correlation is weaker with $r_{\rm S}=0.47,
0.52$ and $0.68$ at $\sim55, 110$ and 165 pc spatial scales, respectively.
However, we note that the 0.32-GHz non-thermal emission is limited by noise.
Deeper observations are required to study the true span and nature of the
correlation at lower frequencies. At 6.2 GHz, the noise is not a limitation and
the correlation can be studied in greater detail.

On the other hand, the non-thermal radio emission at both frequencies
correlates significantly better with FIR emission at $70\,\mu$m as compared to
the MIR emission. The $r_{\rm S}$ between $I_{\rm 70\mu m}$ and $I_{\rm
6.2GHz}$ lies in the range 0.83 and 0.89 for the three spatial scales probed in
our study. At 0.32 GHz, the correlation with the $70\,\mu$m emission is
significantly stronger than with $24\,\mu$m, and $r_{\rm S}$ increases from
0.62 at 55 pc scale to 0.79 at 165 pc scale.

We fitted the data with the form $I_{\rm nt, \nu} = a\, I_{\rm IR}^b$ in
$\log-\log$ space ($\log\, I_{\rm nt, \nu} = b \times \log\, I_{\rm IR} +
\log\, a$) using the ordinary least-square bisector method \citep{isobe90}.
Separate fits were performed for each of the spatial scales, and the best fits
are shown as dash-dotted lines in Fig.~\ref{radioFIR}. In
Table~\ref{corrtable}, we summarize the results of the radio--infrared
relations. The parameter $\sigma_{\rm IR}$ in Table~\ref{corrtable} is a
measure of the scatter or conversely the ``tightness'' of the relations and is
defined as the dispersion around the fit to the data. Since,
the absolute value of the dispersion would depend on the choice of units of the
non-thermal radio and infrared intensities, we have normalized each axis by its
median value before computing $\sigma_{\rm IR}$. One common feature is that
the relation on all spatial scales probed in our study shows stronger
correlation at 6.2 GHz with up to a factor of 1.6 lower
dispersion around the median compared to that at 0.32 GHz. The
dispersion decreases systematically at both the radio frequencies with
increasing spatial scales which is due to the fact that
small-scale fluctuations are smoothed out on large-scales.

Furthermore, we find that the dispersion around the median of the radio--FIR
relation is lower by more than $\sim30$ per cent as compared to that of the
radio--MIR relation. This indicates that the colder dust emitting at $70\,\mu$m
correlates better with non-thermal radio emission as compared to warmer dust
emitting at $24\,\mu$m. Nevertheless, our result is consistent with
\citet{heese14}, who pointed out that the $24\,\mu$m emission arises from dust
heated by star formation activity. Hence, a hybrid indicator of star formation,
i.e., $24\,\mu$m with far-ultraviolet or H$\alpha$ correlates better with the
radio continuum emission as compared to monochromatic emission at $24\,\mu$m.

The slope of the relation between 0.32 GHz and infrared emission at both IR
wavelengths are observed to decrease from roughly linear to sub-linear with
increasing spatial scales. However, within the errors and the noise limitation
of the 0.32 GHz non-thermal emission, this trend is inconclusive. On the other
hand, at 6.2 GHz, within the errors, the slope of the relation with $24$
emission, remains roughly similar with a value of $\sim0.85$ for all three
spatial scales. The slope with the $70\,\mu$m emission is found to be slightly
steeper. The slope of the correlation between FIR and non-thermal emission at
both the radio frequencies are similar within $1.7\sigma$, on all three scales.

\section{Discussion}

\subsection{The galaxy-integrated spectrum}

The total intensity radio continuum spectrum of IC\,10 between 0.32 and 24.5
GHz is consistent with a power-law with $\alpha = -0.34\pm0.01$. The rather
flat shape of this integrated spectrum is due to a high contribution from
thermal free--free emission.  After subtracting the thermal emission, the
non-thermal spectrum is found to be consistent with a power-law with $\ant =
-0.55\pm0.04$. A slight indication of the non-thermal spectrum being curved is
also present. The overall difference between $\alpha$ and $\ant$ is consistent
with what was observed in spatially resolved galaxies at $\sim1$ kpc scales in
high thermal fraction regions \citep{basu12}.

A cursory look at Fig.~\ref{spind} immediately shows the limited 
usefulness of galaxy-integrated spectra or spectra that are spatially
resolved but averaged over $\sim0.5-1$ kpc, a typical linear scale probed by
the currently available telescopes in nearby star-forming galaxies. Our study
of IC\,10 on $\sim50$ pc spatial scale, comparable to the sizes of GMCs and
star-burst regions, shows the importance of aiming for a spatial resolution
that is matched to the relevant physical processes.

Observations at a spatial resolution of 50\,pc such as presented here on IC\,10
show that the galaxy-integrated spectrum is due to a combination of
emission from a non-thermal bubble, optically thick and thin free--free
emission from H{\sc ii} regions, non-thermal emission from recently injected
CRE and diffuse emission from older CRE as one moves away from the disc. It is
therefore difficult to relate the spectral index values measured on kilo-parsec
scales or larger with one particular physical process, as pointed out by
\citet{basu15a} for normal star-forming galaxies. Nearby dwarf irregular
galaxies offer an opportunity to make a comparative study of various emission
mechanisms.

\subsection{``Tightness'' of the radio--FIR relation: a conspiracy} \label{tight}
 
\begin{figure}
\begin{centering}
\begin{tabular}{c}
{\mbox{\includegraphics[width=8.0cm,trim=0mm 12mm 0mm 5mm, clip]{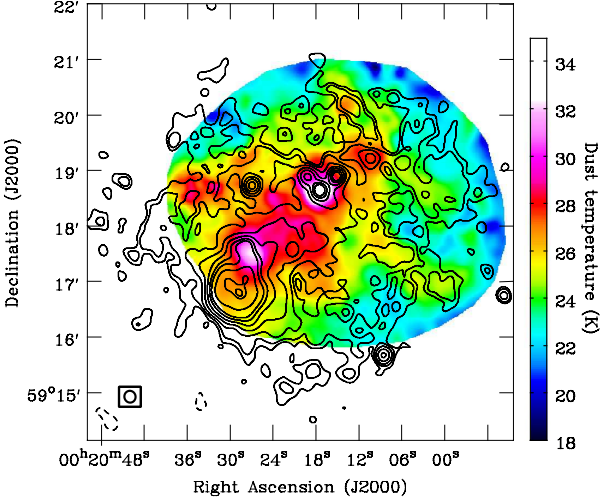}}}
\end{tabular}
\end{centering}
\caption{Distribution of $T_{\rm dust}$ in IC\,10 obtained by fitting a
modified Planck spectrum between 70 and $160\,\mu$m.}
\label{tdust}
\end{figure}

An interesting feature of our study is that, the relation
between non-thermal emission at both 0.32 and 6.2 GHz with that of the FIR
emission shows similar slopes (within $1.7\sigma$) and dispersions (differing
by $\lesssim50$ per cent) on all the three spatial scales probed in our study.
This is surprising because propagation of CREs and scattering
of ultraviolet (UV) photons would result in mixing of different populations
originating from different star formation regions in IC\,10.
Effects of CRE propagation is expected to be larger for lower
energy CREs emitting at 0.32 GHz, which could lead to the slope of the
radio--FIR relation being flatter when studied at 0.32 GHz as compared to that
at a higher frequency such as 6.2 GHz \citep[see][]{basu12a, tabat13}.

The ratio of the flux density at FIR wavelength ($\lambda_{\rm FIR}$) to that
of the non-thermal emission at a radio frequency $\nu$, $I_{\lambda_{\rm
FIR}}/I_{\rm nt, \nu}$, known as the `$q$' parameter, is often used to study
the dispersion of the radio--FIR relation \citep[see e.g.,][]{yun01, apple04,
iviso10a, iviso10b, basu15b}, although, for a non-linear slope of the
radio--FIR relation, $q$ depends on the radio flux densities and is not
suitable for quantifying the correlation \citep{basu15b}. However, it can be
easily shown that $q$ is related to $B_{\rm tot}$, $\ant$ and the dust
temperature ($T_{\rm dust}$) as, 
\begin{equation} \label{eqq}
q\equiv \frac{I_{\lambda_{\rm FIR}}}{I_{\rm nt, \nu}} \propto \left( \frac{n_{\rm UV}}{n_{\rm
CRE}}\right)\,\left(\frac{B_{\lambda_{\rm FIR}}(T_{\rm dust})\, Q(\lambda, a)}{B_{\rm tot}^{(1-\alpha_{\rm nt})}} \right), 
\end{equation}
provided both the FIR and radio emission originate from the same volume. Here,
$n_{\rm UV}$ and $n_{\rm CRE}$ are the number densities of UV photons and CREs,
respectively. $B_{\lambda_{\rm FIR}}(T_{\rm dust})$ is the Planck
function\footnote{Note that, due to lack of standard notation, magnetic field
and Planck function have the same symbol $B$. We denote the Planck function
with a subscript of wavelength ($B_\lambda$) throughout, to distinguish from the
magnetic field strength ($B_{\rm tot}$).} and $Q(\lambda_{\rm FIR}, a)$ is the
wavelength dependent absorption coefficient for dust grains of radius $a$.
Assuming similar dust grain properties and optically thin dust emission
throughout IC\,10, $Q \propto \lambda_{\rm FIR}^{-\beta}$, where we adopt a
typical value for the dust emissivity index $\beta = 2$ \citep{drain84}.

\begin{figure*}
\begin{centering}
\begin{tabular}{cc}
{\mbox{\includegraphics[width=8cm]{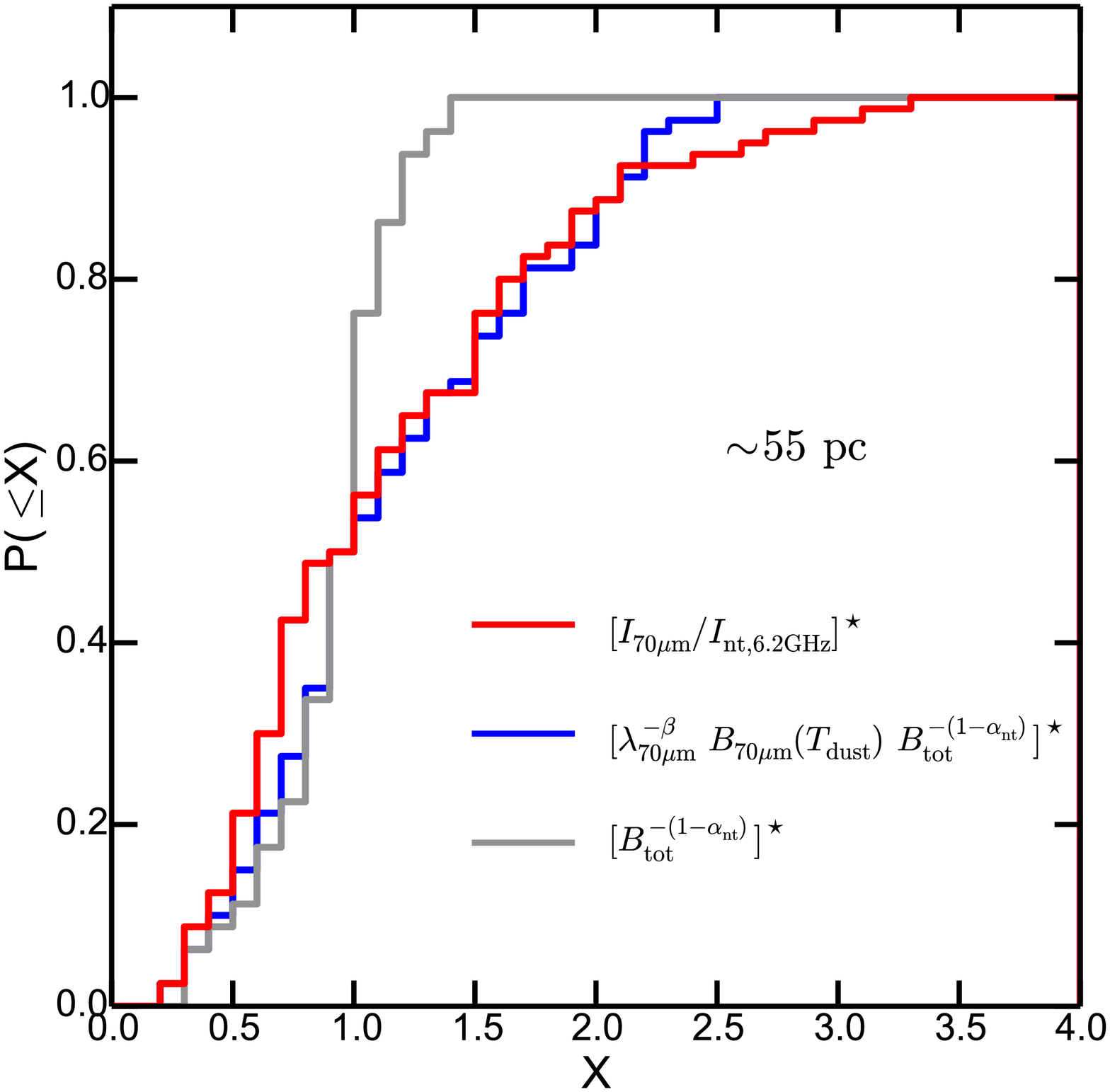}}}&
{\mbox{\includegraphics[width=8cm]{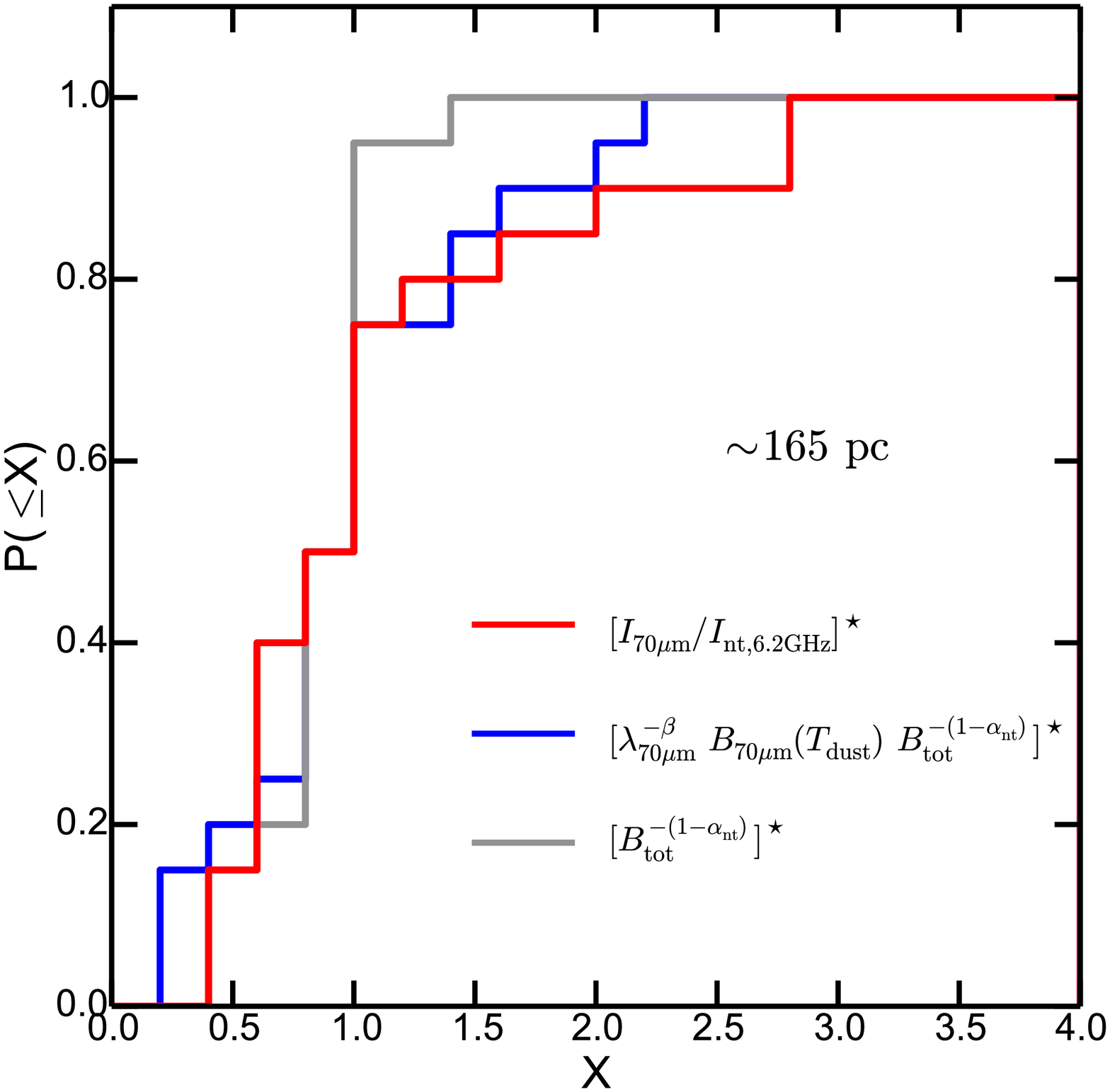}}}\\
{\mbox{\includegraphics[width=8cm]{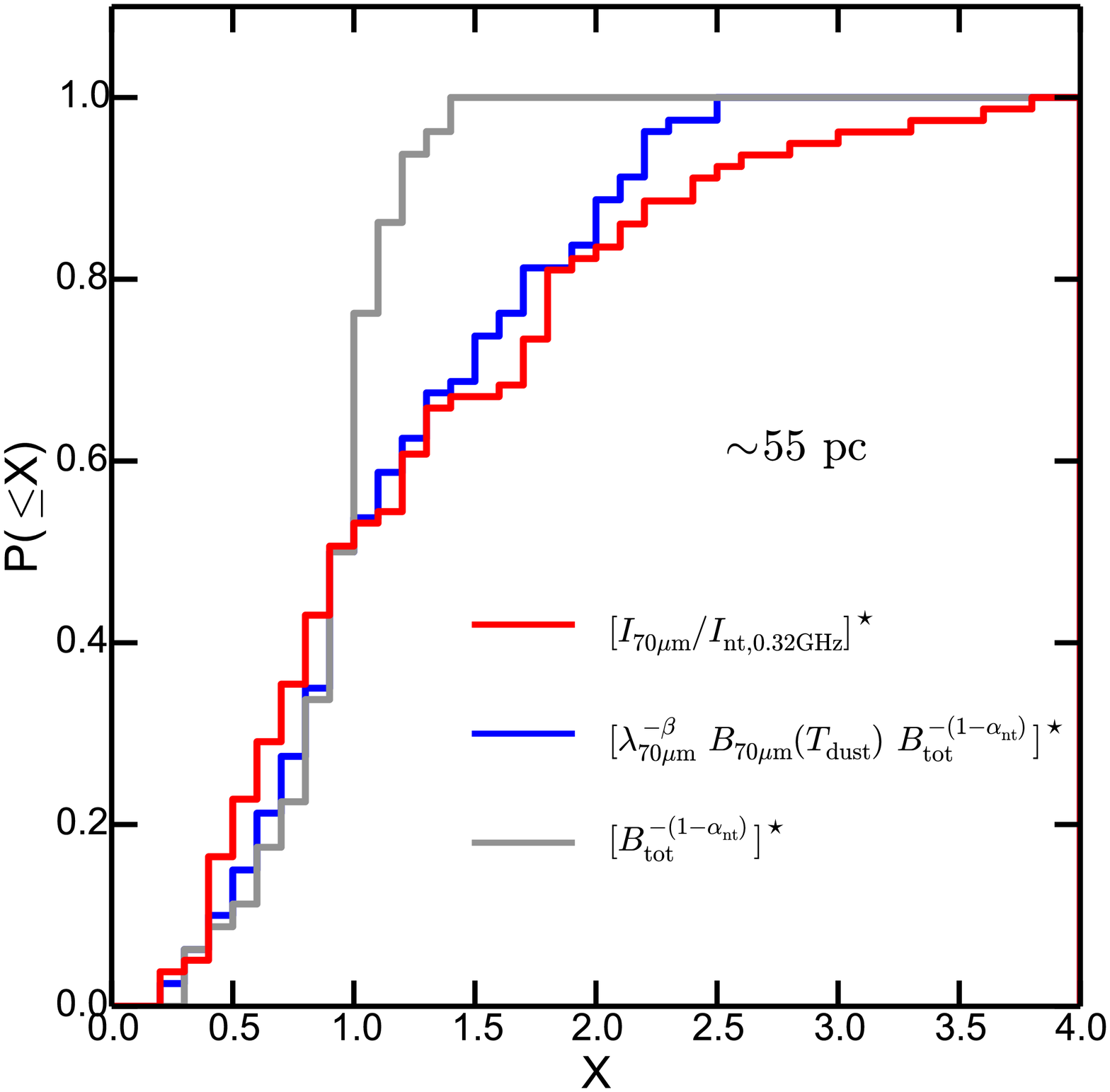}}}&
{\mbox{\includegraphics[width=8cm]{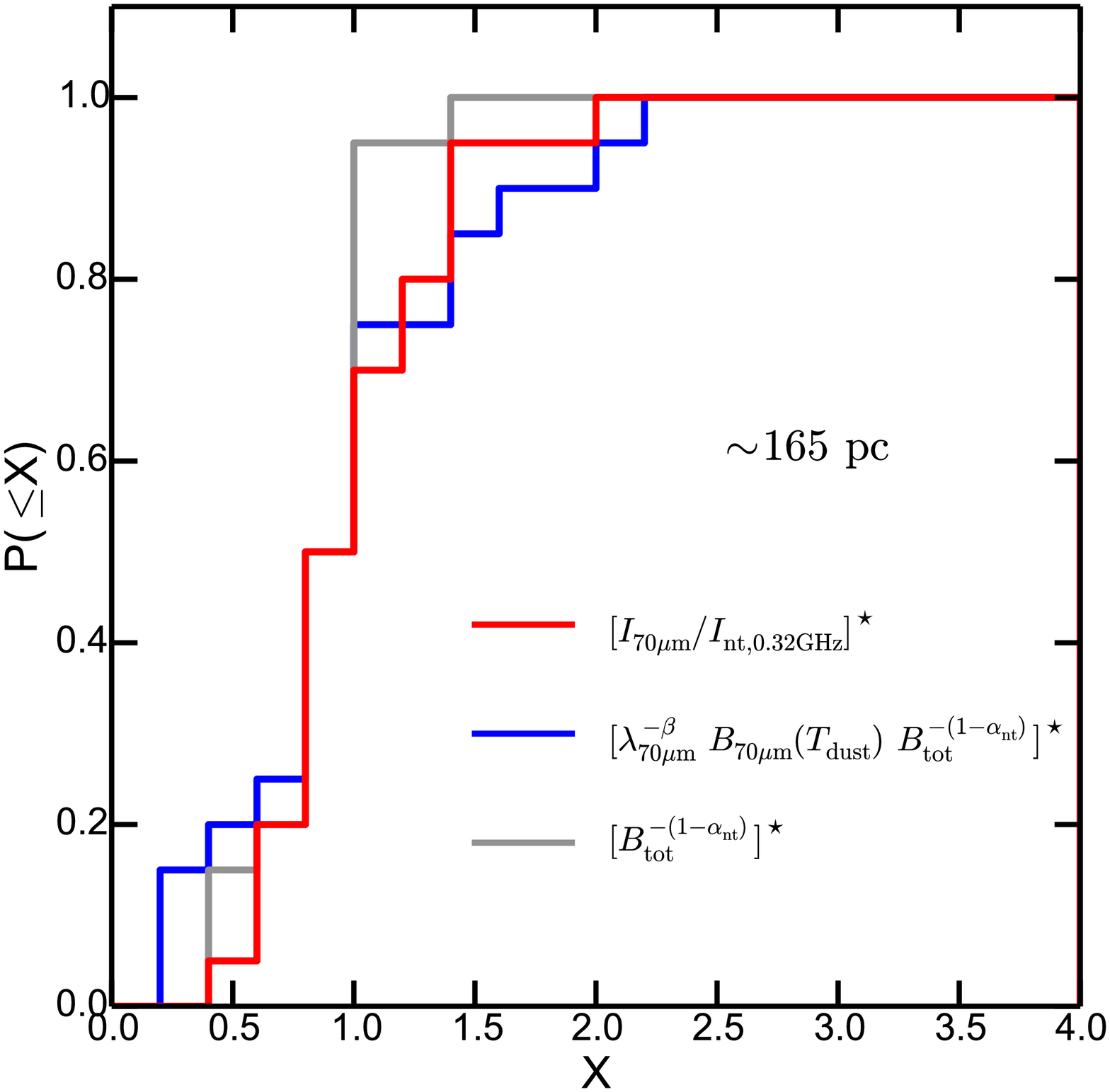}}}\\
\end{tabular}
\end{centering}
\caption{Cumulative distribution function of $X = I_{\rm 70\mu m}/I_{\rm
nt, \nu}$ (shown in red), $X = \lambda_{\rm 70\mu m}^{-\beta}\, B_{\rm \lambda_{\rm
70\mu m}}(T_{\rm dust})\,B_{\rm tot}^{-(1-\ant)}$ (shown in blue) and $X = B_{\rm
eq}^{-(1-\ant)}$ (shown in grey) for $\nu=6.2$ GHz (top panels) and $\nu=0.32$
GHz (bottom panels). The left-hand plots are computed for an aperture of 15
arcsec, i.e., averaged over spatial scale of $\sim55$ pc, while the right-hand
plots are averaged over 45 arcsec apertures corresponding to $\sim165$ pc. The 
$[...]^\star$ indicates the quantities are normalized to their median values, i.e.,
$[X]^\star = X/{\rm median} (X)$.}
\label{cumulative70}
\end{figure*}

Following \citet{tabat07b}, we determine an indicative $T_{\rm dust}$ by
fitting a modified Planck spectrum to the cold dust emission between 70 and
$160\,\mu$m on a pixel-by-pixel basis using {\it Herschel} PACS FIR maps. In
Fig.~\ref{tdust}, we present the $T_{\rm dust}$ map of IC\,10 at a resolution
of 15 arcsec. We find the mean value of $T_{\rm dust}$ to be $\sim25$ K and it
varies widely within IC\,10 having values $\sim20$ K in the outer parts and
reaching up to $\sim35$ K in the H{\sc ii} regions. $T_{\rm dust}$ in IC\,10
and its variation is larger than what is observed in normal star-forming
galaxies ($T_{\rm dust}$ in the range 18--25 K; \citealt{tabat07a, basu12,
kirkp14}) and is consistent with what is observed in low-metallicity dwarf
galaxies ($T_{\rm dust}$ in the range 21--98 K; \citealt{remy13, madde16}).

To understand the origin of the scatter of the radio--FIR correlation, in
Fig.~\ref{cumulative70} we plot the median normalized cumulative distribution
function of $I_{\rm 70\mu m}/I_{\rm nt, \nu}$ (red lines) and compare it to the
median normalized quantities $\lambda_{\rm 70\mu m}^{-\beta}\, B_{\rm 70\mu
m}(T_{\rm dust}) \, B_{\rm tot}^{-(1-\ant)}$ (blue lines) and $B_{\rm
tot}^{-(1-\ant)}$ (grey lines). The top and bottom panels are for non-thermal
emission at 6.2 and 0.32 GHz, while the left and right sides are averaged over
55 and 165 pc spatial scales, respectively. In the figure, we indicate the
corresponding median normalized quantities as $[I_{\rm 70\mu m}/I_{\rm nt,
\nu}]^\star$, $[\lambda_{\rm 70\mu m}^{-\beta}\, B_{\rm 70\mu m}(T_{\rm dust})
\, B_{\rm tot}^{-(1-\ant)}]^\star$ and $[B_{\rm tot}^{-(1-\ant)}]^\star$.
Clearly, the fluctuations of $B_{\rm tot}$ alone are insufficient to produce
the dispersion observed for the correlation between the non-thermal radio
emission and cold dust emission at $70\,\mu$m.  However, on all the scales
probed in our study, the combined fluctuations of $T_{\rm dust}$ and $B_{\rm
tot}$ can well reproduce the distribution of $[I_{\rm 70\mu m}/I_{\rm nt,
\nu}]^\star$. 

\begin{figure}
\begin{centering}
\begin{tabular}{c}
{\mbox{\includegraphics[width=8cm, trim=0mm 0mm 0mm 0mm,clip]{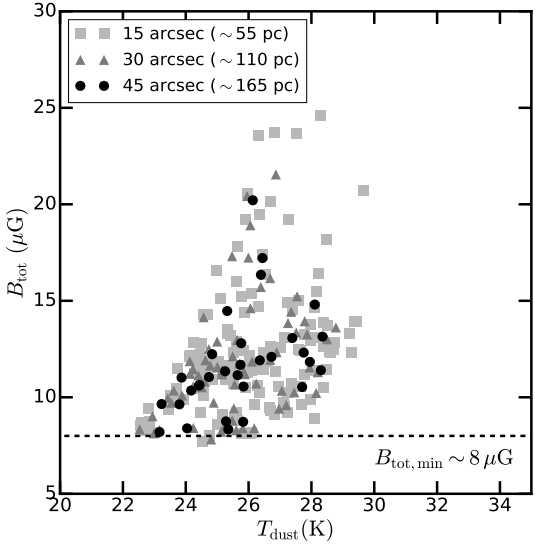}}}
\end{tabular}
\end{centering} 
\caption{Total magnetic field ($B_{\rm tot}$) as a function of dust temperature
($T_{\rm dust}$) in IC\,10. Squares, triangles and circles represents averaging
over 55, 110 and 165 pc spatial scales. The dashed line represents the lower
limit on magnetic field strength detectable by our observations at $3\sigma$
level.} 
\label{bfield_tdust}
\end{figure}

On 165 pc scale, the fluctuations of $B_{\rm tot}$ reproduce the dispersion
around the median $[I_{\rm 70\mu m}/I_{\rm nt, \nu}]^\star$ better, compared to
its fluctuations on smaller scales, as the variations of $T_{\rm dust}$ are
then lower. This is similar to what is observed on kiloparsec scales in normal
star-forming galaxies. At such scales, the variations of $T_{\rm dust}$ are
small and variations in $B_{\rm tot}$ alone are sufficient to reproduce the
dispersion in the radio--FIR relation \citep{basu13}. Hence, our results reveal
an important fact that the tightness of the radio--FIR arises due to a
conspiracy between magnetic field and dust temperature variations on small
scales, while on larger scales, magnetic fields and its coupling with ISM
parameters is responsible. This is evident from Fig.~\ref{bfield_tdust} where
we find the magnetic field strength and dust temperature to show mild
correlation on smaller scales, while on larger scales the correlation is
weaker. This is manifestation of the fact that dust temperature increases with
star formation rate \citep{magne14} and the local magnetic field is related to
the local star formation rate (see Section~\ref{b_sfr}).

Since, the CREs emitting at 0.32 GHz typically have propagation scale-length
$\sim2-4$ times larger than those emitting at 6.2 GHz, depending on whether
they are transported by diffusion or advection at Alfv{\'e}n speed, one would
expect the dispersion of the radio--FIR relation to be significantly larger at
0.32 GHz. However, in contrary, we find that the distribution of the radio--FIR
relations at both the radio frequencies are already well reproduced by invoking
the variations of both $B_{\rm tot}$ and $T_{\rm dust}$, on scales $\gtrsim55$
pc. This suggests, as per Eq.~\ref{eqq}, that the fluctuations of $n_{\rm
UV}/n_{\rm CRE}$ are small and thus, both CRE transport length ($l_{\rm CRE}$)
and the mean-free path of UV photons ($l_{\rm mfp}$) are smaller than $\sim55$
pc. 

In a typical ISM, $l_{\rm mfp}$ is $\sim20-100$ pc, roughly the size of the
Str{\"o}mgren sphere ionized by OB stars \citep{oster06}. Hence, scattering of
UV photons are unlikely to be significant. On the other hand, since the
magnetic fields in the disc of IC\,10 are predominantly tangled
\citep{chyzy16}, the CREs are expected to be transported via the streaming
instability at the Alfv{\'e}n speed, $V_{\rm A} = B_{\rm tot}/\sqrt{4\,\pi\,
\rho_{\rm gas}} \sim 12~\rm km\,s^{-1}$. The gas density $\rho_{\rm gas} \sim
8.6\times10^{-24} ~\rm g\,cm^{-3}$ was estimated from a H{\sc i} surface
density map of IC\,10.\footnote{IC\,10 was observed as a part of the LITTLE
THINGS \citep{hunter12}. To compute $\rho_{\rm gas}$, we integrated the H{\sc
i} surface density map within the $3\sigma$ contour of the radio emitting disc
and assumed a scale-height of $700$ pc. The data was downloaded from:
https://science.nrao.edu/science/surveys/littlethings.} Therefore, for $l_{\rm
CRE}$ to be smaller than $\sim55$ pc, the CREs must have been produced
$\lesssim5$ Myr ago, consistent with the starburst scenario proposed by
\citet{vacca07} and also with the spectral ageing analysis by \citet{heese15}.
The young population of CREs in IC\,10 gives rise to $\ant \sim -0.5$ observed
within the disc.

\subsection{Slope of the radio--FIR relation and $B_{\rm tot}$--$\rho_{\rm
gas}$ relation}

Under the condition of energy equipartition between magnetic field and kinetic
energy of the turbulent gas, MHD simulations reveal that the magnetic field
($B_{\rm tot}$) and gas density ($\rho_{\rm gas}$) are coupled as $B_{\rm tot}
\propto \rho_{\rm gas}^{\kappa}$, where $\kappa\sim0.4-0.7$ \citep{chand53,
mousc76, fiedl93, cho00, kim01}. Building on the theory first proposed by
\citet{nikla97b}, \citet{dumas11} showed that, in the optically thin regime,
the slope of the radio--FIR relation, $b$, the Kennicutt-Schmidt (KS) power law
index, $n$, connecting the star formation rate and gas surface densities, and
$\ant$ are related to $\kappa$ as:
\begin{equation} 
\kappa = \frac{(n + 1)\,b}{(3-\ant)}.
\end{equation}
This relation is valid provided both radio and FIR emission originate from the
same emitting volume. In the previous section we argued, in IC\,10 the
non-thermal radio emission at 6.2 GHz and the FIR emission at $70\,\mu$m are
correlated due to coupling of magnetic fields with the ISM on $\sim165$ pc
scale.  Therefore, we use the slope of the radio--FIR relation $b=1.08\pm0.11$,
on 165 pc scale and, the mean value and dispersion of the observed
$\ant=-0.62\pm0.15$. In dwarf irregular galaxies such as IC\,10, the gas
density is dominated by atomic H{\sc i} and the galaxy-averaged KS relation's
index $n$ is found to be $0.91^{+0.23}_{-0.25}$ \citep{roych14}. However, to
compare with our spatially resolved study, we determine the index of the
spatially resolved KS relation for IC\,10 as $1.15\pm 0.06$.  Using these
values we estimate $\kappa=0.64\pm0.17$ which is consistent with numerical
simulations of a turbulent ISM. This indicates that the assumption of energy
equipartition between magnetic field and cosmic ray particles is valid on 165
pc scales.  Such a relation was confirmed in normal star-forming spiral galaxies
at a spatial scale of $\sim1$ kpc, i.e., the diffusion scale-length of CREs
emitting at 1.4 GHz \citep{basu12a}. For the first time we confirm this
relation in a dwarf irregular galaxy on sub-kpc scales.

\subsection{Magnetic field and star formation: $B_{\rm tot}$--$\Sigma_{\rm
SFR}$ relation} \label{b_sfr}

\begin{figure}
\begin{centering}
\begin{tabular}{c}
{\mbox{\includegraphics[width=8cm, trim=0mm 0mm 0mm 0mm,clip]{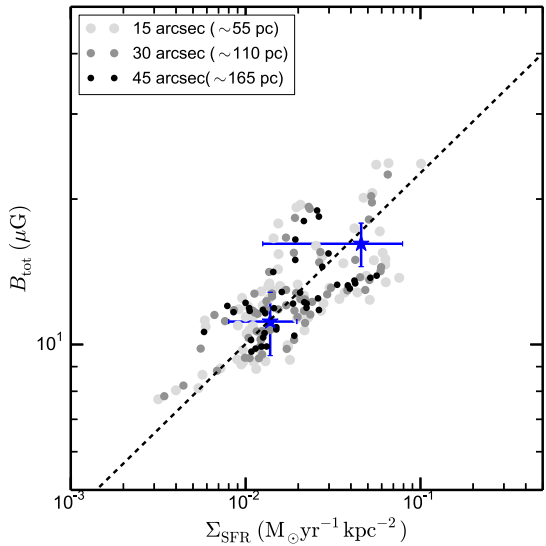}}}
\end{tabular}
\end{centering} 
\caption{Total magnetic field ($B_{\rm tot}$) as a function of surface star
formation rate ($\Sigma_{\rm SFR}$) in IC\,10. The statistical
error on the data points are smaller than the scatter of the plot. However, a
systematic error up to factor of 2 for SFR calibration can be present. The blue
stars are $\Sigma_{\rm SFR}$ determined on 165 pc scales and are binned in
$B_{\rm tot}$ such that the total $\rm SFR$ in each bin is $\geq \rm 10^{-2.5}
\, M_\odot\,yr^{-1}$. The dashed line is the bisector fit to the data points
averaged over 15 arcsec in $\log-\log$ space. The best-fit power-law is found
to be $(51\pm6)\, {\Sigma_{\rm SFR}^{0.35\pm0.03}}$.} 
\label{bfield_sfr}
\end{figure}

Recently, in a semi-analytical model to explain the correlation between radio
and FIR emission, \citet{schle13, schle16} pointed out that the correlation is
mainly driven by the coupling between magnetic fields and surface density of
star formation rate ($\Sigma_{\rm SFR}$) of the form $B_{\rm tot} \propto
\Sigma_{\rm SFR}^{1/3}$. The coupling is established due to star-formation
driven turbulent amplification of the magnetic field through the fluctuation
dynamo operating on small-scales ($\lesssim1$ kpc). In Fig.~\ref{bfield_sfr},
we plot the spatially resolved total magnetic field strength as a function of
the $\Sigma_{\rm SFR}$ in IC\,10. We used the H$\alpha$ map, as explained in
Section~\ref{sfr}, to create a map of $\Sigma_{\rm SFR}$.  We find the magnetic
field strength to be strongly correlated with $\Sigma_{\rm SFR}$ having
$r_S=0.7$ on 55 pc scales.  Using the bisector method fit in $\log$--$\log$
space we find,
\begin{equation}
\left(\frac{B_{\rm tot}}{\rm \mu G}\right) = (51\pm6) \left(\frac{\Sigma_{\rm
SFR}}{\rm M_\odot\,yr^{-1}\,kpc^{-2}}\right)^{0.35\pm0.03}. \label{b-sfr}
\end{equation}
The power-law index is similar to what is expected for
star-brust driven turbulent amplification of the magnetic field $B_{\rm tot}$
and is likely responsible for establishing the radio--FIR correlation on
sub-kpc scales \citep{schle13, schle16}.

By comparing simulation results of synthesized stellar populations
\citep{dasil12} with that of SFR indicators, \citet{dasil14} found that, for
$\rm SFR \lesssim 10^{-2.5}\, M_\odot\, yr^{-1}$ the H$\alpha$-based SFR
indicators can be strongly affected by stochasticity. Thus on small scales
and/or in regions of low star formation, the usual star formation rate
calibrations can suffer from stochasticity due to insufficient sampling of the
IMF, thereby giving rise to biases at $\gtrsim0.5$ dex level \citep{dasil14}.
Hence, although the statistical error on the $\Sigma_{\rm SFR}$ are smaller
than the scatter in Fig.~\ref{bfield_sfr}, there can be significant errors on
the H$\alpha$-flux to SFR conversion factor.

To check the effects of unreliable SFR conversion and incomplete IMF sampling
issues on small scales, we have binned the $\Sigma_{\rm SFR}$ determined on 165
pc scales\footnote{We have made sure that none of the pixels in the
$\Sigma_{\rm SFR}$ map are counted multiple times by ensuring that there are no
overlapping pixels within adjacent regions.} in $B_{\rm tot}$ such that, in
each bin the total $\rm SFR \geq 10^{-2.5}\, M_\odot\,yr^{-1}$. The binned data
are shown as blue stars in Fig.~\ref{bfield_sfr} which closely follows the
$B_{\rm tot}$--$\Sigma_{\rm SFR}$ relation determined on smaller scales. The
error on $B_{\rm tot}$ represents the bin-size and error on $\Sigma_{\rm SFR}$
is computed by adding the statistical error in quadrature to a factor of 2
error for the conversion of H$\alpha$ flux to SFR.

Such $B_{\rm tot}-\Sigma_{\rm SFR}$ relation has been reported using the
galaxy-averaged $\Sigma_{\rm SFR}$ and magnetic field strengths in a sample of
dwarf galaxies where the power-law indices were found to be $0.30\pm0.04$
\citep{chyzy11} and $0.25\pm0.02$ \citep{jurus14}. In fact, the $B_{\rm
tot}$--$\Sigma_{\rm SFR}$ relation is also observed to hold well for
star-forming spiral galaxies with a similar power-law index, both globally and
locally \citep{nikla97b, heese14}.

\subsection{On the normalization of $B_{\rm tot}$--$\Sigma_{\rm SFR}$ relation}

The total magnetic field strength of IC\,10 is similar to that observed in
large spiral galaxies. \citet{chyzy16} argued that magnetic field amplification
driven by the classical large-scale $\alpha$--$\Omega$ dynamo is insufficient
to produce the strong fields observed in IC\,10.  Here, we explore the effects
of magnetic field amplification in turbulent plasmas on small-scales. Our study
shows that the power-law index of the $B_{\rm tot}$--$\Sigma_{\rm SFR}$
relation agrees well with the predictions of semi-analytical models of
turbulent field amplification and with other galaxy-averaged observations.
However, the normalization factor ($B_0$) of $51\pm6~\mu$G is about a factor of
2 higher than what is expected. For example, \citet{schle13} predict the
normalization to be $\sim26~\mu$G using typical values for the physical
parameters of the ISM.\footnote{Note that, \citet{schle13} reported the
normalization factor to be $\sim12~\mu$G in units of $0.1~ \rm
M_\odot\,yr^{-1}\,kpc^{-2}$.} They showed that the normalization depends on ISM
parameters as, 
\begin{equation}
B_0 \approx \langle\rho\rangle^{1/6} \left( C/\tilde{C} \right)^{1/3} (8\,\pi\,f_{\rm sat})^{1/2}. \label{bnorm}
\end{equation}
Here, $\langle\rho\rangle$ is the gas density, $f_{\rm sat}$ is the fraction of
turbulent kinetic energy converted to magnetic energy for a saturated dynamo
and,  $C$ and $\tilde{C}$ describe the injection rate of turbulent supernova
energy and the normalization of the KS relation, respectively.  The average density
of atomic hydrogen in IC\,10, $\langle\rho_{\rm HI}\rangle \approx
8.6\times10^{-25}~\rm g\,cm^{-3}$, is similar to the density of $10^{-24}~\rm
g\,cm^{-3}$ assumed by \citet{schle13} and thus insufficient to explain the
difference, because $B_0$ depends only weakly on the density.

In a detailed study of the KS relation in dwarf irregular galaxies, where the
surface gas density is dominated by atomic hydrogen, like in IC\,10,
\citet{roych14} found $\log\,\tilde{C}_{\rm dwarf}=-3.84$, which is different
from large star-forming disc galaxies ($\log\,\tilde{C}_{\rm disc}=-3.6$;
\citealt{kenni98}). $\tilde{C}$ is $\sim1.7$ times lower in dwarf galaxies than
in spiral galaxies which however will increase $B_0$ only by a factor of
$\sim1.2$.

The constant $C$ in Equation~\ref{bnorm} is given as $C = \nu_{\rm SN}\,
\epsilon\, E_{\rm SN}$ \citep{schle13}. Here, $\nu_{\rm SN}$ is the rate of
core-collapse supernovae, $\epsilon$ is the fraction of supernova energy
($E_{\rm SN}\sim10^{51}$ erg) deposited as turbulent energy.  Numerical
simulations of diffusive shock acceleration in supernova remnants shows
$\epsilon\sim0.05$ \citep{tatis08, bellAR13, bellAR14}. Now, the rate of
supernovae is given by $\nu_{\rm SN}= \left(f_{\rm M}/\langle M_{\rm
SN}\rangle\right) \times \Sigma_{\rm SFR}$. Here, $f_{\rm M}$ is the
mass-fraction of stars resulting in core-collapse supernovae, typically stars
with mass $\gtrsim8\,\rm M_\odot$, and $\langle M_{\rm SN}\rangle$ is the
average mass per supernova. \citet{schle13} used $f_{\rm M}\sim 0.08/{\rm
M_\odot}$ assuming a standard Kroupa-type IMF \citep[][Schleicher, priv.
comm.]{kroup01}. However, in low metallicity environments of dwarf irregular
galaxies like IC\,10, the IMF can be top heavy, following a Salpeter-type IMF,
but with a flatter index skewed towards high mass stars \citep{nakam01,elmeg06,
oey11}. Therefore, assuming a power-law type IMF, $N(M) \propto M^{-\beta}\,
dM$, with a slope $\beta=2$, we estimate $f_{\rm M} \approx 0.14/{\rm
M_\odot}$.  Thus, the combined effect of $\left(C/\tilde{C}\right)^{1/3}$ will
give rise to an increase up to a factor of at most $\sim1.4$ of the
normalization $B_0$.

\citet{schle13} assumed $f_{\rm sat}\approx5$ per cent to derive the
normalization factor $B_0\approx26~\mu$G. In our case, $f_{\rm sat}\approx10$
per cent is required to explain the observed normalization of
$B_0\approx50~\mu$G in IC\,10. In fact, even after accounting for a factor of 2
systematic error on the SFR calibration, $f_{\rm sat}\gtrsim5$ per cent is
required to explain the observed value of $B_0$. This indicates that the
starburst driven turbulent dynamo in IC\,10 is highly efficient in converting
the turbulent kinetic energy into magnetic energy. This is perhaps the reason
for the relatively strong magnetic field strengths in IC 10.

It is interesting that the efficiency of the small-scale dynamo in IC\,10 is
$\gtrsim 5$ per cent. For the case of compressively driven turbulence (which is
relevant in case of driving by supernovae), MHD simulations suggests that
$f_{\rm sat}$ decreases with increasing Mach number, and drops significantly
below 5 per cent for Mach numbers $>1$ \citep[see Fig. 3 of][]{feder11}. In
fact, for highly compressible turbulence, the theoretical saturation level lies
in the range 0.13--2.4 per cent \citep{schob15}. At the transonic point,
compressible turbulence has a maximum efficiency of $\sim3$ per cent. Thus,
compressible turbulence is less likely to be the driver of the small-scale
fluctuation dynamo in IC\,10.

On the other hand, the small-scale dynamo is more efficiently excited by
solenoidal forcing as it can produce tangled field configurations filling a
larger volume \citep{feder11}. Therefore, a possible way to achieve $f_{\rm
sat} \gtrsim 5$ per cent is that the turbulence is driven by mildly supersonic
solenoidal forcing with Mach numbers $\sim$2--10. Another scenario is that we
are observing a strong magnetic field in the aftermath of a star burst, which
happened a few Myr ago. The star formation in IC\,10 subsided 1 Myr ago
\citep{heese15}, whereas the advection time scale is $\sim 10^7$ yr
\citep{chyzy16}, so that the magnetic field has decayed only a little since the
end of the star burst. Therefore, a detailed MHD simulation is necessary to
understand the nature of magnetic field amplification in star-burst dwarf
galaxies.

\subsection{Implications for tracing star formation at high redshifts}

Studying the cosmic evolution of star-formation history is fundamental to
understanding the physical properties of the ISM in late-type galaxies that we
observe in the nearby universe.  However, to probe star formation in
cosmologically distant galaxies via tracers like UV and H$\alpha$ emission
suffers from obscuration, both internal and along the line of sights making
surveys at these wavelengths incomplete \citep[see][for a review]{madau14}.
Therefore, infrared emission is used to trace SFR at high redshifts, well
before the peak epoch of cosmic star formation history \citep[][]{magne09,
karim11, magne14}. However, the sources in the deep imaging performed with the
current infrared telescopes, such as the {\it Herschel} and {\it Spitzer} are
sufficiently confused \citep{jarvi15} and evolution of dust temperatures can
lead to biases \citep{smith14, basu15b}. Of late, one uses the advantage of the
radio--FIR relation to use the radio continuum as a proxy to infer SFR in high
redshift galaxies \citep{seymo08, smolc09}. This is one of the major drivers of
science with the next generation radio facilities such as the {\it Square
Kilometre Array} \citep[see e.g.,][]{jarvi15}.

Star-burst dwarf galaxies are believed to contribute significantly to the
co-moving star formation rate leading up to the peak in
cosmic star formation around redshifts of 2 \citep{buitr13, alavi16, ribei16}.
Our study of IC\,10 suggests the same principle governs the radio--FIR
relation in star-burst dwarf galaxies as in large star-forming galaxies and
hence their radio emission can be used to trace star formation in the early
universe.  Using an independent study with CO and radio emission,
\citet{leroy05} reached the same conclusion regarding the universality of the
interdependent relations.  However, the high contribution of the thermal
component to the radio emission needs to be taken into account while interpreting the
results, especially in the intermediate observed-frame frequencies between 1.4
and 10 GHz.  This would otherwise increase the scatter of the derived
radio--SFR relation leading to systematic biases \citep{galvi16}.  At
frequencies $\gtrsim10$ GHz, the thermal free--free emission will dominate
which is a direct tracer of star-formation \citep[see e.g.,][]{murph15}.  On
the other hand, at lower frequencies ($\lesssim 1$ GHz), the correlation could
suffer due to free--free absorption.  Therefore, multi-radio frequency
observations are necessary to separate these effects. A similar precaution was
suggested in a semi-analytic study of the radio--FIR relation by
\citet{schob16}.

This study reveals the significant role $T_{\rm dust}$ can play in
shaping the radio--FIR relation on different spatial scales. In fact, $T_{\rm
dust}$ differs for different galaxy types and with star formation rates
\citep{hwang10, magne14}. Hence, for studying the relation at higher redshifts
where different galaxy types contribute to the co-moving star formation rate
density, systematic changes in the dust temperature will lead to biases.
Further, currently available models in the literature explaining the origin of
the radio--FIR relation ignore the variations of $T_{\rm dust}$. It is clear
from our study of IC\,10 that different physical mechanisms are at play on
different scales to maintain the correlation. A more careful treatment of the
$T_{\rm dust}$ variations with metallicity and SFR must be considered in order
to interpret the physics of the correlation.

\section{Summary}

We have studied the dwarf starburst galaxy, IC\,10, at 0.32 GHz using the GMRT.
We achieved a map rms noise of $\sim150~\mu$Jy beam$^{-1}$ and an angular
resolution of $13\times 12$ arcsec$^2$, making this the highest sensitivity and
angular resolution image of IC\,10 available in the literature below 1.4 GHz.
We summarize our main findings in this section.

\begin{enumerate}[(i)]

\item At 0.32 GHz, the radio continuum emission from IC\,10 originates from
complex intrinsic structures and bright H{\sc ii} regions. The radio continuum
emission closely follows the H$\alpha$ emission suggesting it to be a good
tracer of star formation.
 
\item We have estimated the thermal free--free emission from IC\,10 using
H$\alpha$ emission as the tracer. We estimate the thermal fraction to be
$\sim0.2$ at 0.32 GHz and $\sim0.53$ at 6.2 GHz. This is significantly higher
than what is observed in star-forming spiral galaxies. In fact, we find that
the radio continuum emission from the compact H{\sc ii} regions visible in the
H$\alpha$ map to be 100 per cent thermal in origin.

\item Several of the compact H{\sc ii} regions show evidence of thermal free--free
absorption in our 0.32 GHz observations. Using three frequency radio continuum
spectra, in two of the H{\sc ii} regions we constrain the thermal electron
densities to be $\sim11$ and $22~\rm cm^{-3}$. Higher angular resolution and
lower radio frequency observations with LOFAR are necessary to study the
nature of the thermal emission from the compact H{\sc ii} regions.

\item We detect the non-thermal bubble towards the south-eastern edge of IC\,10
which has a projected size of $\sim300\times 100$ pc$^2$. The thermal emission
is lowest in this bubble with only 3.5 per cent of the total radio emission at
0.32 GHz being thermal in origin. The bubble is bound by giant molecular clouds
detected in CO($J = 1\to0$) observations, their magnetic fields possibly
keeping the CREs confined.

\item We estimate the average equipartition magnetic field strength of
$\sim10~\mu$G in IC\,10, similar to that of normal spiral galaxies.  The field
is strongest in the non-thermal bubble with an average field of $\sim20~\mu$G.

\item In IC\,10, the non-thermal radio emission is well
correlated with MIR emission at $24\,\mu$m and FIR emission at $70\,\mu$m on
spatial scales of 55, 110 and 165 pc. On all scales, the FIR emission better
correlates with the non-thermal radio emission, and results in a significantly
tighter relation than one with the MIR emission.

\item On small scales, the dispersion of the radio--FIR
relation is caused by the combined effect of variations in magnetic field and
dust temperature. While, on larger scales, the fluctuations in dust temperature
are smoothed out and the dispersion mainly arises from magnetic fields
variations. The radio--MIR relation originates directly from star formation. 

\item The total magnetic field is strongly correlated with star formation as
$B_{\rm tot} \propto {\rm SFR}^{0.35\pm0.03}$. This is in good agreement with
what is expected for star formation driven amplification of magnetic fields via
the fluctuation dynamo on small scales.

\item The efficiency of the turbulent dynamo in IC\,10 is $\gtrsim5$ per cent
suggesting that supernova driven compressible turbulence is unlikely to be the
driver of small-scale magnetic field amplification.

\end{enumerate}

\section*{Acknowledgments}

We thank Dominik Schleicher for fruitful discussions on properties of the
small-scale dynamos and Deidre Hunter for help with the H$\alpha$ data. We
thank Sui Ann Mao for careful reading of the manuscript and useful comments.
We thank the anonymous referee for helpful suggestions and
constructive review of the manuscript. We also thank the staff of the GMRT
that made these observations possible. GMRT is run by the National Centre for
Radio Astrophysics of the Tata Institute of Fundamental Research. This work is
based (in part) on observations made with the {\it Spitzer Space Telescope},
which is operated by the Jet Propulsion Laboratory, California Institute of
Technology under a contract with NASA. This research has made use of the
NASA/IPAC Infrared Science Archive (IRSA) and NASA/IPAC Extragalactic Database
(NED), which are operated by the Jet Propulsion Laboratory, California
Institute of Technology, under contract with the National Aeronautics and Space
Administration. This paper is partly based on observations with the 100-m
telescope of the MPIfR (Max-Planck-Institut f{\"u}r Radioastronomie) at
Effelsberg. JW acknowledges support from the UK Science and Technology
Facilities Council [grant number ST/M503514/1]. EB and LH acknowledge support
from the UK Science and Technology Facilities Council [grant number
ST/M001008/1].

\bibliographystyle{mnras}

\bibliography{abasu_etal_IC10.bbl}

\bsp

\label{lastpage}

\end{document}